\documentclass{aa}  

\usepackage{graphicx}
\usepackage{float}
\usepackage{txfonts}
\usepackage{amsmath,amssymb}
\usepackage{longtable}
\usepackage{booktabs}
\usepackage{soul}
\usepackage{amsmath,amssymb}

\usepackage[colorlinks,allcolors=blue,breaklinks=true]{hyperref}
\usepackage{xcolor}

\begin{document}

   \title{Exploring the variability of young stars with Gaia DR3 light curves}

   \author{C. Mas\inst{1}\thanks{These authors contributed equally to this work.}
          \and
          J. Roquette\inst{1}$^{\star}$
          \and
          M. Audard\inst{1}
          \and
          M. Madarász\inst{2,3,4}
                    \and
          G. Marton\inst{2,3}
                    \and
          D. Hernandez\inst{5}          
          \and    
          I. Gezer \inst{2,3}
          \and
          O. Dionatos \inst{6}
          }

    \institute{Department of Astronomy, University of Geneva, Chemin Pegasi 51, 1290 Versoix, Switzerland
               \and
            Konkoly Observatory, Research Centre for Astronomy and Earth Sciences, Hungarian Research Network (HUN-REN), Konkoly Thege Mikl\'{o}s \'{U}t 15-17, 1121 Budapest, Hungary
            \and
            CSFK, MTA Centre of Excellence, Budapest, Konkoly Thege Miklós út 15-17., H-1121, Hungary   
            \and 
             Department of Experimental Physics, Institute  of Physics, University of Szeged, D{\'o}m t{\'e}r 9, 6720 Szeged, Hungary
            \and            
            Universität Wien, Institut für Astrophysik, T\"urkenschanzstrasse 17, 1180 Wien, Austria
             \and
             Natural History Museum Vienna, Burgring 7, 1010 Vienna, Austria
             }

   \date{}

  \abstract
   {Photometric variability is a defining characteristic of young stellar objects (YSOs) that can be traced back to a range of physical processes that occur at different stages in the formation and early evolution of young stars. The Gaia third Data Release (GDR3) has provided an unprecedented dataset of photometric time series, including 79\,375 light curves for sources classified as YSO candidates. Through its all-sky coverage, Gaia provides a unique opportunity for large-scale studies of YSO variability.}{Our goal was to characterise the GDR3 sample of YSO variables to better identify the recurrence of YSO variability modes (caused by accretion, extinction, rotation modulation, etc.).
   We made a pilot study of the applicability of the asymmetry (M) and periodicity (Q) variability metrics to characterise YSO variability with Gaia light curves. By adapting the Q-M metrics for sparse and long-term light curves, we sought to bridge the gap between low- and high-cadence survey insights on YSO variability.}{We adapted the Q-M method for Gaia.
Through a refined sample selection, we identified sources with an appropriate sampling for the Q-M method. We used the generalised Lomb Scargle periodogram and structure functions to derive characteristic variability timescales.}{We successfully derived Q-M indices for ~23\,000 sources in the GDR3 YSO sample. These variables were then classified into eight variability morphological classes. We linked the morphological classes with physical mechanisms using H$\alpha$ as a proxy of accretion and $\alpha_\mathrm{IR}$ indices to gauge whether circumstellar material was present.}{We demonstrate that the Q-M metrics can be successfully applied to study the sparse time series of Gaia. We applied it successfully to distinguish between the various variability modes of YSOs. While our results are generally consistent with previous high-cadence short-term studies, the long GDR3 time span yields a larger variety of variability mechanisms.}


   \maketitle
       \nolinenumbers
%

\section{Introduction}

Gaia \citep{Gaia2016, Gaia2023j} is a cornerstone mission by ESA that is in operation since mid-2014. In 2022, the Gaia Data Release 3 (GDR3) provided photometric and astrometric observations for 1.8 billion stars. GDR3 included epoch photometry covering 34 months of the mission provided for 11\,754\,237 variable sources \citep[stellar and extragalactic;][]{Eyer2023} and all sources in the field of view (FOV) of the Gaia Andromeda Photometric Survey \citep[GAPS][]{Evans23}. The variables were classified into 25 classes with the GDR3 general variability detection path via supervised machine learning \citep{Rimoldini2023}, and 79\,375 of them were classified as young stellar objects (YSOs). This sample of YSO candidates was validated by \citet{Marton2023} by cross-matching it with existing YSO catalogues from optical and infrared surveys. This revealed that 40\,000 sources were newly identified YSOs. In this paper, we further explore the GDR3 YSO variable sample by studying their variability modes in more detail. 

Variability is a prevalent characteristic of YSOs, which are observed throughout their entire electromagnetic spectrum \citep{Joy1945,Herbst94,Stelzer2007A&A...468..463S,Carpenter2001,MoralesCalderon2011ApJ...733...50M,Venuti2015A&A...581A..66V,Roquette2020A&A...640A.128R}. This variability can be traced back to a wide range of physical mechanisms that occur in the formation and evolution of young stars, including mass accretion, star-disc interaction, disc dissipation, and outflows \citep[e.g.][]{Fischer23}. Understanding YSO variability is thus fundamental to advancing our knowledge of how stars form and evolve during their first million years. 

Attempts to devise variability classification schemes capable of grasping the assortment of observed YSO light curves have been made since the 1990s \citep[e.g.][]{Herbst94}. Although light curve characteristics such as amplitude, timescale, and periodicity were recurrent ingredients, these earlier classification schemes were limited to the specific data for which they were designed.  
Based on an initial sample of 162 YSOs in the young cluster NGC 2264, \citet{Cody_14} proposed a new morphology classification scheme for YSO light curves. This scheme stems from two variability metrics designed to quantify the level of asymmetry (M index) and periodicity (Q index) in the light curves and allow for sorting sources into seven variability classes: strictly periodic (P), eclipsing binary (EB), quasi-periodic symmetric (QPS), quasi-periodic dippers (QPD), aperiodic dipper (APD), burster (B), and stochastic (S). Two more classes, multi-periodic and long trend (L), were added to include sources with characteristics that were missed by the default Q-M classification. 
Previous research comparing Q-M classification results with proxies of young stellar evolution (e.g. disc IR excess and accretion-induced H$\alpha$ emission) helped to associate these classes with various physical mechanisms behind the variability of YSOs. A scenario emerged according to which P light curves are attributed to magnetically induced cold spots on the stellar surface viewed from high-inclination angles \citep[e.g.][]{Cody_2022}. Higher-amplitude QPS light curves are associated with accretion-induced hot spots at the stellar surface \citep[e.g.][]{Venuti2015A&A...581A..66V,Cody_2018}. Dipper variability is linked to the AA Tauri phenomenon \citep[e.g.][]{Bouvier1999A&A...349..619B,Alencar2010}, where magnetic star-disc-interaction is thought to lift dusty disc material above the disc mid-plane, which leads to periodic or aperiodic occultation. Bursting variables have been correlated with observation of intense ongoing accretion, suggesting a common origin \citep[e.g.][]{Stauffer2014AJ....147...83S}. S variability is thought to originate from accretion-related processes combined with low viewing angles \citep{Cody_2018}. While the Q-M classification scheme was initially designed with the aim to apply it to high-cadence space-based light curves from the Convection, Rotation and planetary Transits satellite \citep[CoRoT][]{CoRoT2009A&A...506..411A} and Spitzer telescope, it has since been employed in the context of varied epoch surveys including other high-cadence surveys such as the Kepler mission K2 \citep{Cody_2018,Venuti_2021,Cody_2022}, but also ground-based surveys such as the Zwicky Transient Facility \citep[ZTF;][]{Hillenbrand2022,Wang2023,JiangHillenbrand2024}, the All-Sky Automated Survey for Supernovae \citep[ASAS-SN;][]{Bredall2020}, and the Vista Variables in the Vía Láctea Extended \citep[VVVX;][]{2024MNRAS.533..841O}.

We present the results of a pilot study that addressed the applicability of the Q-M metrics to GDR3 light curves. 
It is challenging to apply this method because the observational window resulting from the Gaia scanning law is sparse and unevenly spaced. Most sources have only 30 to 100 epoch observations that span 900 days, which requires a full adaptation of the original method by \citet{Cody_14}, based on 40-day-long high-cadence light curves with more than 4\,000 epochs. Still, Gaia offers all-sky coverage, which is in contrast to the modest sample of a few hundred YSOs that were studied by high-cadence surveys focused on a handful of nearby star-forming regions. Beyond the caveats of its sparse sampling, Gaia therefore provides the opportunity to derive general statistics of YSO variability from a more extensive and diverse sample. Additionally, the longer time coverage of Gaia allowed us to probe longer variability timescales, which are relevant for our understanding of the scope and prevalence of YSO variability mechanisms. 

This paper is organised as follows. Sect.~\ref{sec:data} describes the data we used. With the goal of testing whether the Q-M indices are applicable to GDR3 data, we performed several tests that we describe in Sect.~\ref{section_preprocessing} to identify and discard sources or epoch observations that might be affected by instrumental biases. The clean sample was divided into two variability levels, and the highest-variability sample was used to benchmark our method. In Sect.~\ref{section_method} we describe our implementation of the Q-M indices for GDR3, with timescales derived using generalised Lomb-Scargle periodograms or structure functions. We also discuss the cases without reliable timescale or those where the timescale is too long compared to the full light curve duration.
Sect.~\ref{section_results} presents the results for the Q-M indices, timescales, and morphological classes. We also look into the Q-M indices in relation to the disc properties and evolutionary stage of our samples, and we use the $\alpha_\mathrm{IR}$ index and the H$\alpha$ emission line. In Sect.~\ref{section_discussion} we discuss our results and present example light curves from each class, and we discuss the typical timescale, amplitude, and physical mechanisms behind them. We further discuss the limitations of our methods and compare our results to previous high-cadence and ground-based studies. Finally, we summarise our findings in Sect.~\ref{section_conclusion}.

\section{Data}\label{sec:data} 

   \begin{figure}[thb]
   \centering
      \includegraphics[width=1\linewidth]{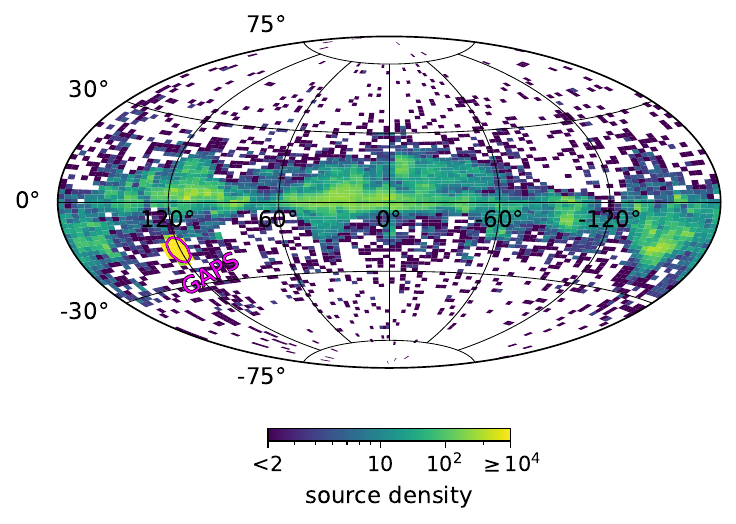}
      \includegraphics[width=1\linewidth]{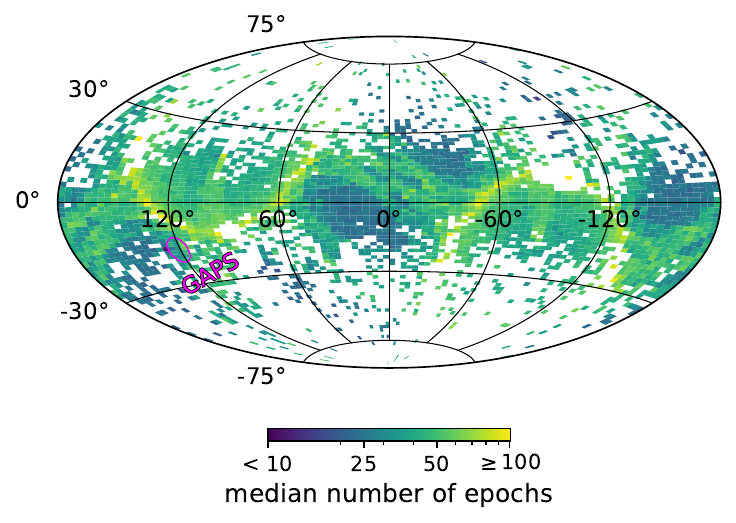}
      \caption{Sky distribution of sources with GDR3 epoch data (YSO and GAPS samples). The location of the GAPS field is highlighted by an ellipse and a label in magenta. The distributions are shown in Galactic coordinates and in an Aitoff projection. In the top panel, colours indicate the source density, and in the bottom, colours reflect the average number of epochs for the light curves given their sky position.
 }
         \label{fig:sky_distributions}
   \end{figure}

\subsection{Epoch data}

GDR3 epoch data cover 34 months of Gaia operation from July 25, 2014 to May 28, 2017. We used two epoch photometry datasets from GDR3:

\begin{itemize}
    \item YSO sample: composed of 79\,375 variable sources identified as YSO candidates as part of the GDR3 variability processing \citep{Eyer2023,Rimoldini2023,Marton2023}.
    \item GAPS: The Gaia Andromeda Photometric Survey \citep[GAPS][]{Evans23} released epoch photometry for over 1.2 million sources located in a region of 5.5$\degr$ radius centred on Andromeda. GAPS included the release of epoch photometry for all sources, i.e. both variables and non-variables, showcasing the full power of the Gaia epoch data.
\end{itemize}

Because of the Gaia scanning law, the number of transits per source highly depends on the sky position. This is illustrated in Fig. \ref{fig:sky_distributions}. For the YSO sample, the number of epochs ranged from 14 to 138, and from 5 to 110 for GAPS. On average, YSO light curves have more epoch observations than GAPS sources. General source properties (mean magnitude, coordinates, etc.) were retrieved from the  \citet{GaiaArchive}. Light curves were retrieved using the \texttt{astroquery.gaia} package \citep{AstroqueryGinsburg2019AJ....157...98G}. For this study, we focus our analysis on the $G$-band epoch data with the goal of classifying variability types. Analysis of other bands ($G_{BP}$, $G_{RP}$) may help distinguish between specific YSO processes, but are beyond our current scope.

\subsection{Supplementary data}

In Sect.~\ref{sec:spurious_var}, we use data from GDR3 spurious variability table \citep[\texttt{vari\_spurious\_signals;}][]{Holl2023} to identify sources affected by scan-angle spurious variability. This data was retrieved directly from the  \citet{GaiaArchive}.

\subsubsection{Epoch data from the literature}

In Sect.~\ref{section_discussion}, we compare our results with previous literature investigating YSO variability using K2 and ZTF light curves. In \citet{Cody_2022}, K2 light curves were employed to study sources in the Taurus star-forming region, with 61 sources in common with our YSO sample; light curves were provided to us through private communication with the authors. \citet{Hillenbrand2022}, \citet{Wang2023}, and \citet{JiangHillenbrand2024} used ZTF light curves to study YSO variability in the North America and Pelican Nebulae Complex, the Perseus Molecular Cloud, and the Mon R2 region, respectively. These studies included 509 sources in our YSO sample. ZTF light curves for these sources were downloaded from the ZTF archive \citep{Masci2019PASP..131a8003M} through the ZTF data access tool at the NASA/IPAC Infrared Science Archive. The time span of these light curves was trimmed to match the data release used in these previous studies, and the pre-processing of light curves followed the steps described in these studies. 

\subsubsection{YSO properties}\label{sec:YSOproperties}

To support validation of our results in Sect.~\ref{section_discussion}, we matched our sources with catalogues that provide measurements related to the youth and evolutionary stages of YSOs. For measurements related to the $H\alpha$ line, a proxy of accretion, we used:

\begin{itemize}
\item The NEMESIS (New Evolutionary Model for Early Stages of Stars with Intelligent Systems) Catalogue of YSOs for the Orion star formation complex \citep{roquette2024} compiles youth indicators for sources in that region, including $H\alpha$ measurements for 5,231 objects in our YSO sample.

\item GALAH DR4 
\citep[Galactic Archaeology with HERMES Survey Data Release 4;][]{Buder2024arXiv240919858B} includes EW($H\alpha$)  from high-resolution spectroscopy for 2\,085 of our sources. 
\item LAMOST DR10 {v2.0} \citep[The General Survey of the Large Sky Area Multi-Object Spectroscopic Telescope Data;][]{Zhao2012RAA....12..723Z} includes EW($H\alpha$) measured from low-resolution spectra for 2\,172 sources.

\item The Gaia-ESO Survey \citep{Gilmore2022AA...666A.120G, Randich2022AA...666A.121R} DR5 5.1 catalogue \citep{Hourihane2023AA...676A.129H} provides EW($H\alpha$) and full widths at 10\% of the peak flux measurements from high-resolution spectra for 751 and 765 sources, respectively. 

\item Gaia DR3 includes spectroscopic data products based on low-resolution spectra from the blue and red photometers \citep[][]{BPRPspectra2023AA...674A...2D}, providing measurements of EW($H\alpha$) with pseudo-equivalent widths estimated as part of the Extended Stellar Parametriser for Emission-Line Stars \citep[ESP-ELS][]{Apsis2023AA...674A..26C} for 56\,964 sources.
\end{itemize}

\section{Data pre-processing} \label{section_preprocessing}

  \begin{figure}[htb]
   \centering
      \includegraphics[width=0.9\linewidth]{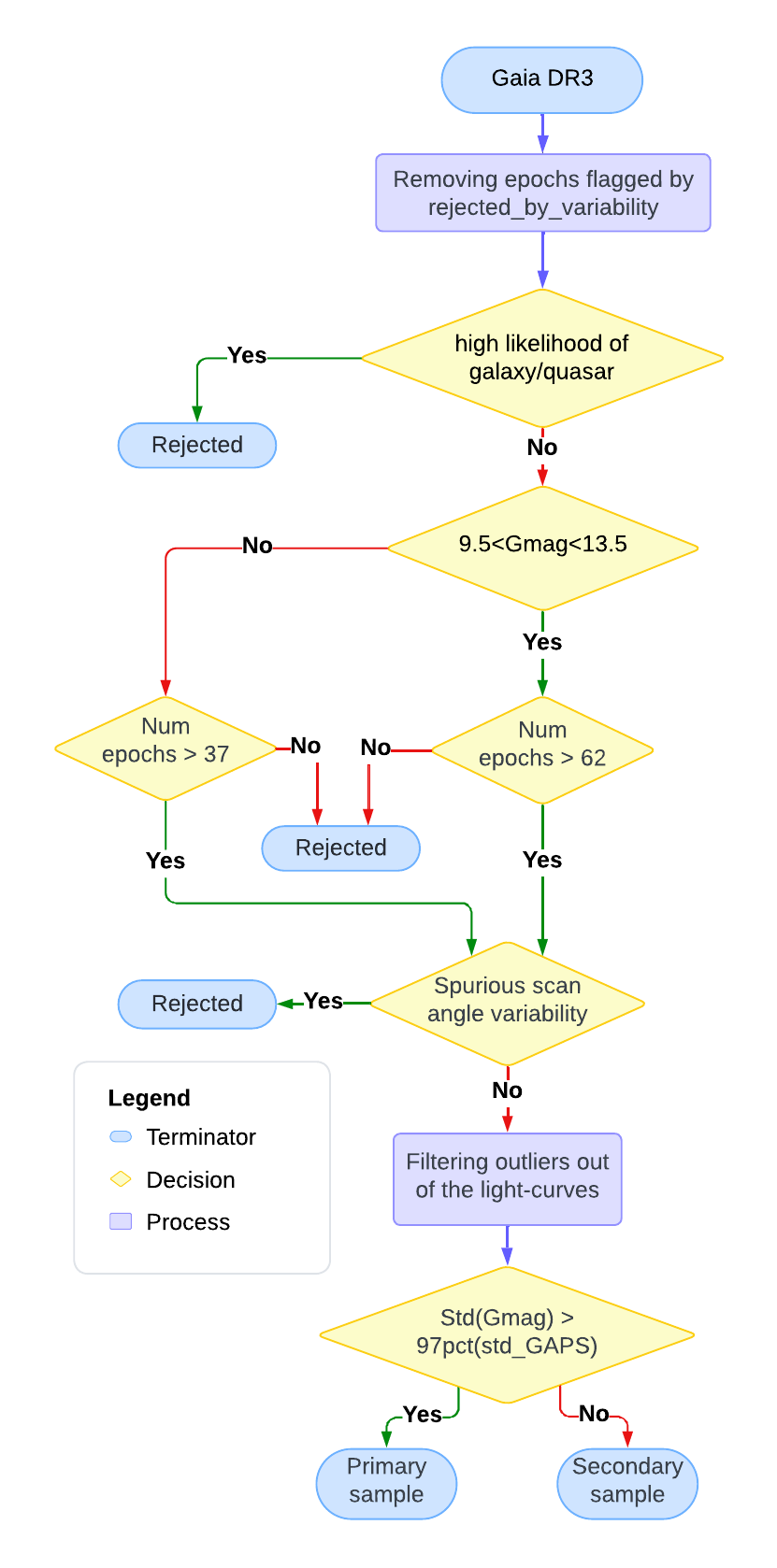}
      \caption{Workflow of filtering and cleaning steps to remove extragalactic contamination prior to asymmetry and periodicity analysis (Sect.~\ref{sec:extragalactic_filtering}), sources with too few epochs (Sect.~\ref{sec:numb_epochs}), affected by a spurious scan-angle (Sect.~\ref{sec:spurious_var}), outliers (Sect.~\ref{sec:spurious_outliers}), and, finally, sorting the sources based on their degree of variability (Sect.~\ref{sec:selection_high_var}).}
         \label{fig_decision_tree}
   \end{figure}

Fig.~\ref{fig_decision_tree} summarises the pre-processing and filtering steps we employed to ensure that only the most reliable light curves were kept in our sample. Our first cleaning step was to remove likely spurious epoch observations based on GDR3 flag \texttt{rejected\_by\_variability}, which provides information on transits considered as outliers or extreme values and rejected as part of GDR3 variability processing steps \citep[see sect. 3.1 of][]{Eyer2023}.

\subsection{Contamination by extragalactic sources} \label{sec:extragalactic_filtering}

We excluded $14\,722$ sources from the GAPS sample which were flagged with a probability over 80\% of being a galaxy or a quasar, based on the Gaia DR3 classification probabilities \texttt{classprob\_dsc\_combmod\_galaxy} and \texttt{classprob\_dsc\_combmod\_quasar} \citep{GDR3_Extragalactic_2023A&A...674A..41G,Delchambre2023A&A...674A..31D}. Six sources were similarly removed from the YSO sample.

\subsection{Minimum number of epochs}\label{sec:numb_epochs}

In App.~\ref{appendix_bootstrap} we performed tests to determine the minimum number of epochs required in the light curve to ensure statistically robust variability indices. All tests were carried out as a function of magnitude to account for the different levels of photometric noise in the sample. Based on these tests,  we rejected light curves with fewer than 62 observations for $G=9.5-13.5$ mag and 37 for outside this range. This filtering step yield 44\,918 sources in our YSO sample (57\% of the full catalogue) and 524\,096 sources in the GAPS sample (42\%).

\subsection{Spurious variability}\label{sec:spurious_var}

\citet{Holl2023} investigated spurious variability in GDR3 light curves due to scan-angle effects \citep[combination of the Gaia mirror aspect ratio, spin-axis precession, and orbital period,][]{Gaia2016}. Attempts to model and filter out the spurious frequencies caused by these scan-angle effects would require assumptions such as the underlying variability signal having a sinusoidal waveform, which are not reliable priors for YSO variability. Hence, instead of attempting to mask these signals, we followed the Gaia data-processing literature \citep{Holl2023,Distefano2022} to identify and remove sources likely affected by spurious scan-angle signals (see App.~\ref{app:spurioyuus_var}). After this step, we deemed 34\,793 light curves in the YSO sample as reliable for further analysis (44\% of the full catalogue), and 479\,936 in the GAPS sample (39\%).

\subsection{Outlier removal}\label{sec:spurious_outliers}

Filtering large outliers is an important for studying the degree of asymmetry in the light curves, as outliers can bias the M index calculation (see Sect.~\ref{subsection_QM}). When dealing with high-cadence data, \citet{Cody_14} addressed this issue by 5$\sigma$-clipping the light curves. With the densely sampled light curves of CoRoT and K2, this approach was unlikely to remove relevant variability features. However, due to the sparsity of the GDR3 light curves, this type of filtering needs to be addressed more carefully. 
We have done so in App.~\ref{app:outlier}, where 636 sources in our sample had at least one outlier removed. All affected sources still fulfilled the minimum number of epochs requirement after outlier removal.

\subsection{Selection of highly variable sources} \label{sec:selection_high_var}

  \begin{figure}[htb]
   \centering
      \includegraphics[width=1\linewidth]{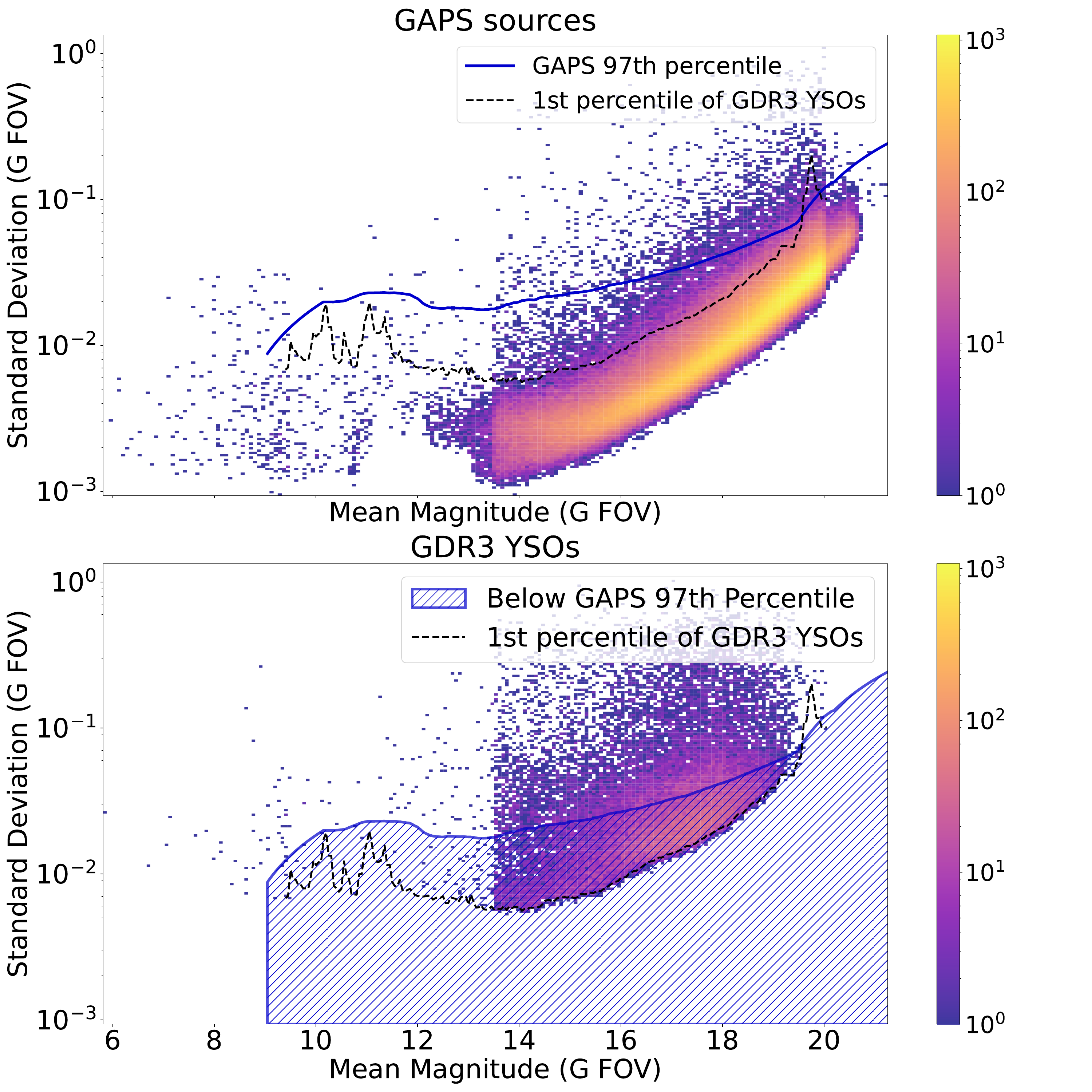}
      \caption{Standard deviation versus $G$ magnitude for light curves in the GAPS (top) and YSO (bottom) samples. The colour bar indicates source counts. The solid blue line marks the 97th percentile for the GAPS sample, and the hatched region identifies YSO sources lying below this percentile (secondary sample), while primary-sample sources lie above it. In both panels, the black dashed line shows the lower envelope (1st rolling percentile) of the YSO distribution to aid visual comparison.}
         \label{fig_std_mag_percentiles}
   \end{figure}

Initial tests revealed that low-amplitude QPS was the most common variability mode in the YSO sample, dominating the light curves of 41\% of the sources analysed (see Sect.~\ref{sec:QPS} and Table~\ref{tab:gdr3_allsky}). However, to assess the applicability of the Q-M indices to GDR3 light curves, we concentrated our pipeline development on a sample selected to include a higher proportion of less common variability modes. This was achieved by focussing our initial analysis on a sub-sample of the most variable sources.

To support the identification of the most variable sources as a function of magnitude, it was relevant to first understand the behaviour of the YSO sample compared to non-variable stars also observed by GDR3. Although GDR3 does not provide light curves for non-variable YSOs, this comparison was still possible based on the GAPS sample, which includes GDR3 light curves for all sources in the Andromeda field, irrespective of their variability status. The GAPS and YSO samples share GDR3 photometric performance, albeit having different sky locations and numbers of transits (see Fig.~\ref{fig:sky_distributions}). A comparison between the two samples as a function of their average G magnitude and standard deviation (std) is provided in Fig.~\ref{fig_std_mag_percentiles}, where we display properties of light curves that passed pre-processing steps described in this section (34\,793 YSO and 479\,334 GAPS light curves), i.e. already accounting for the minimum number of epochs required for stable variability indices (Sect.~\ref{sec:numb_epochs},  App.~\ref{appendix_bootstrap}). The GAPS sample is predominantly made up of non-variables, with roughly 1\% of its sources selected by GDR3 variability processing. This is evident in Fig.~\ref{fig_std_mag_percentiles}, where the bulk of the GAPS distribution (yellow regions, top panel) is located below the lower envelope of the YSO sample distribution (derived as the 1st rolling percentile of that distribution). 

We defined a threshold for selecting the most significant variables as a function of magnitude based on rolling percentiles of the GAPS distribution, which were
estimated by binning the GAPS distribution using $0.05$ mag bins from $G=$9.1 to 21.9 mag. Each magnitude bin was required to contain at least 500 sources, and bin sizes were sometimes increased by a factor of 1.1 until this minimum number of sources was satisfied. Rolling percentiles were then calculated within each bin, and a Savitzky-Golay filter \citep{SavitzkiGolay1964A} was applied to smooth it. The 97$^{th}$-percentile threshold provided the best compromise between selecting highly variable sources while still retaining a large enough sample of variables.

Finally, we selected 10\,739 sources in the YSO sample with a std larger than 97\% of GAPS sources in their magnitude bin. These sources are the primary sample on which we focused the adaptation of the Q-M method (Sect.~\ref{section_method}). The remaining 24\,054 sources below the 97$^{th}$ percentile of the GAPS distribution are hereafter named secondary sample; these were also analysed once the workflow for the Q-M analysis was settled.

\section{Methods}\label{section_method}

\subsection{Asymmetry and periodicity metrics} \label{subsection_QM}

We explored the applicability of the Q-M indices to characterise the GDR3 light curves. The Q index is defined as \citep{Cody_14,Hillenbrand2022}:

\begin{equation}
    Q = \frac{\mathrm{std}^2_{r} - \langle \sigma^2 \rangle}{\mathrm{std}^2_{m} - \langle \sigma^2 \rangle}, \label{Q_eq}
\end{equation}

\noindent where $\mathrm{std}^2_{r}$ is the variance of the residuals of the signal, estimated by subtracting a waveform estimation (derived by smoothing the phase-folded light curve, folded to the characteristic variability timescale) from the raw light curve. $\mathrm{std}^2_{m}$ is the variance of the original light curve, and $\langle \sigma^2 \rangle$ is the square of the typical uncertainty (calculated here as the average of all squared errors associated with the data). The Q index is thus a measure of the ratio of the variance of the residuals to the general variance of the light curve. Lower Q values close to zero indicate high periodicity. In contrast, higher Q values (close to unity) reflect irregular or stochastic variability. The derivation of the Q index assumes that the light curve contains a signal that may repeat at a characteristic timescale. This signal can be modelled by phase-folding the light curve to its characteristic timescale and employing a smoothing technique to model the average waveform of the signal. A residual curve is then calculated as the difference between the raw-light curve and this waveform.

The M index measures the asymmetry of the light curve by comparing how the upper and lower brightness deciles deviate from the median. It is is defined as
\begin{equation}
    M = \frac{\langle m_{10\%} \rangle -m_{50\%}}{\sigma_m}, \label{M_eq}
\end{equation}
where $\langle m_{10\%} \rangle$ is the mean of the highest and lowest 10\% of light-curve points, $m_{50\%}$ is the median magnitude, and $\sigma_m$ is the standard deviation.  When working in magnitudes, light curves with dipping behaviour have $M\geq 0.25$, those with bursts have $M\leq-0.25$, and symmetric light curves have $M\approx0$.

\subsection{Characteristic timescale}\label{sec:timescale}

\begin{figure}[htb]
   \centering
      \includegraphics[width=0.95\linewidth]{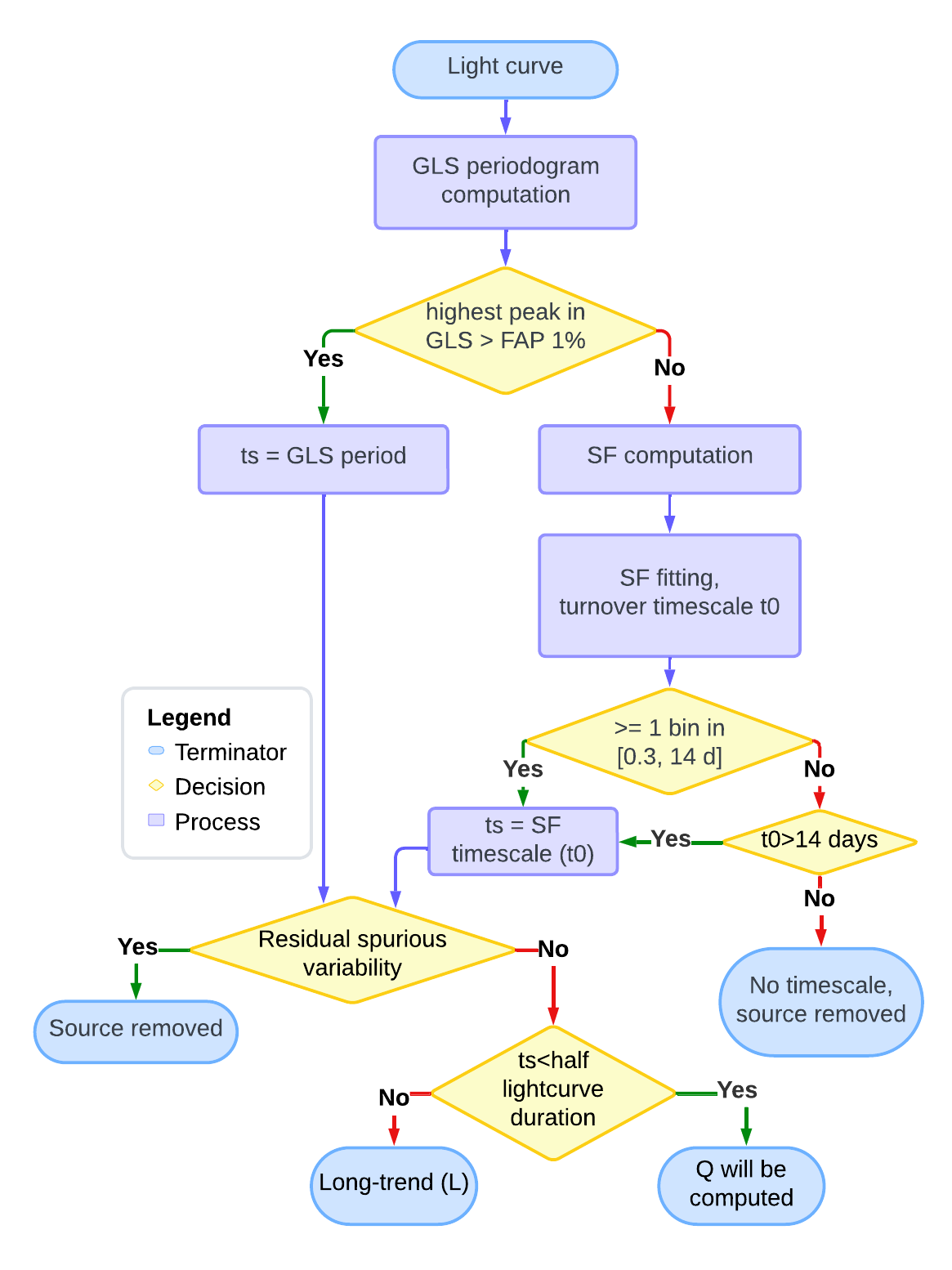}
      \caption{Workflow for deriving the characteristic timescale required for the periodicity analysis with the Q index.}
      \label{figure_process_timescale_finding}
\end{figure}

The capital step to quantify the periodicity with the Q index is to derive a characteristic timescale of variability. Fig. \ref{figure_process_timescale_finding} summarises our workflow for timescale derivation: We first applied the generalised Lomb-Scargle periodogram to identify periodic timescales, assessing their reliability using the false alarm probability (Sect.~\ref{gls_ts}). If no reliable periodic timescale was found, we then computed their structure function to derive an aperiodic timescale (Sect~\ref{sf_ts}). Sources were discarded if neither method provided a trustworthy timescale. A final step removed sources with timescales affected by the typical spurious variability of Gaia (Sect~\ref{sec:residual_spur_var})

\subsubsection{Generalised Lomb-Scargle periodogram}\label{gls_ts}

\begin{figure}
   \centering
      \includegraphics[width=1\linewidth]{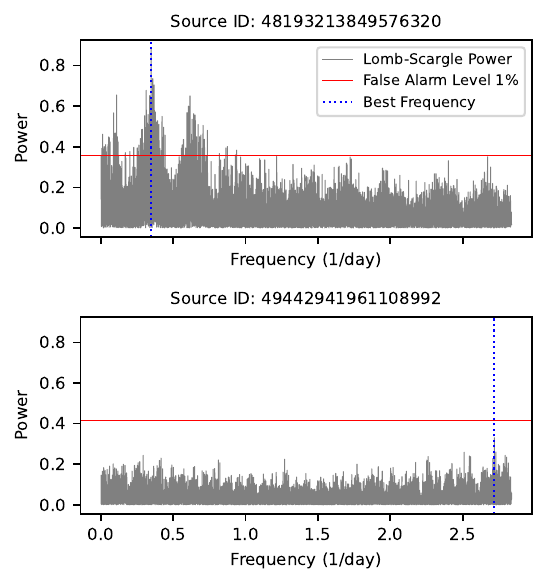}
      \caption{Example GLS periodograms used to derive periodic timescales.The upper panel shows a periodogram with a peak above the 1\% FAL, from which a periodic timescale is obtained, while the lower panel shows a case with no such peak, where an aperiodic timescale is instead derived from the structure function.}
      \label{fig_periodograms}
   \end{figure}

We used the generalised Lomb-Scargle periodogram \citep[GLS;][]{Lomb1976, Scargle1982, Zechmeister2009,Vanderplas2018} to search for periodic variability in our sample. We searched for periods between twice the median time step of the light curve (0.5 days) and the entire light curve duration (900 days), with the frequency of the highest peak recorded as the most representative timescale of the light curve variability. The reliability of GLS timescales was evaluated considering a false-alarm probability level (FAL) of 1\%, with a \citet{Baluev2008MNRAS.385.1279B} FAL estimation. Sources without a peak above the FAL were considered aperiodic. Fig.~\ref{fig_periodograms} shows examples of GLS periodograms for a periodic and an aperiodic source. A total of 24\,067 sources ($\sim69\%$ of the YSO sample) had periodic timescales measured with the GLS. 3\,721 of these timescales were longer than half the total time span of the light curve, and were stored as evidence of L type variability, but were not used to derive the Q index.

\subsubsection{Structure function}\label{sf_ts}

In the previous section, 10\,726 sources did not show any peak above the FAL in their GLS, indicative of periodicity. These sources are likely aperiodic variables. Still, their study with the Q metric required the inference of a characteristic timescale for their aperiodic variability. In their original implementation, \citet{Cody_14} employed the \texttt{PeakFind} \citep{Findeisen_2015ApJ...798...89F} algorithm for aperiodic timescale inference, which requires high-cadence sampling to resolve detailed inflexions in light curves \citep{Cody_2022} and thus is unattainable with sparsely sampled data sets such as GDR3. Alternatively, we inferred aperiodic timescales using structure functions (SF). SFs are widely used in other fields to constrain variability timescales, with varied applications in the study of Active Galactic Nucleus \citep[AGN]{Simonetti85, Kozlowski2016}, other variable stars \citep{1999A&AS..136..421E}, and with a few applications to YSOs albeit in the context of high-cadence surveys\citep[][]{Venuti_2021, Lakeland2022}. In the context of Gaia DR2, structure functions were successfully used to investigate short timescale variables \citep{Roelens2017}. 

 \begin{figure}[htb]
   \centering
      \includegraphics[width=0.95\linewidth]{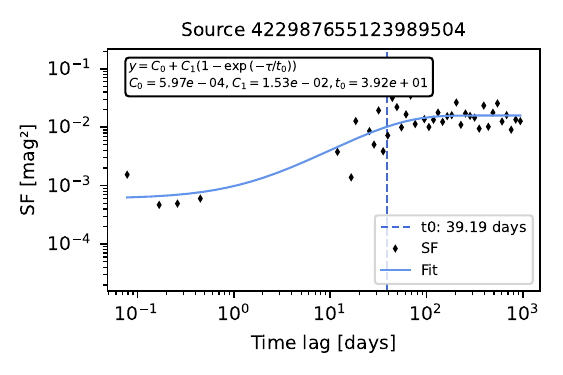}
      \includegraphics[width=0.95\linewidth]{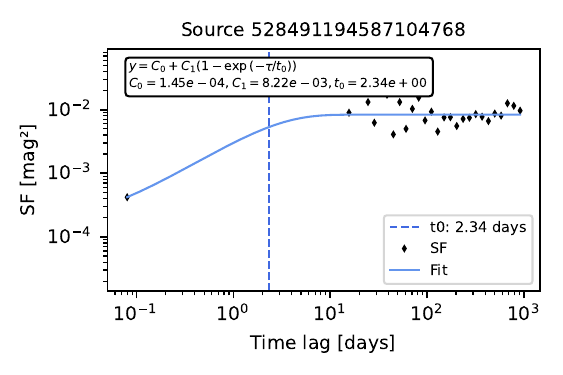}
      \caption{Examples of sources with aperiodic timescale derived from fitting an aperiodic model to their binned SF. The best-fit models are shown in blue. The top panel shows an aperiodic variable with a timescale of 39 days. The lower panel presents an example of a source for which no reliable timescale could be determined, because Gaia’s sparse sampling produces a blind interval in its SF.}
      \label{fig_sf}
   \end{figure}
   
Following \citet{Hughes1992} and  \citet{Roelens2017}, the SF of a light curve is defined as
\begin{equation}
\mathrm{SF}(\tau) = \frac{1}{N(\tau)} \sum_{i>j}{(m_i - m_j)^2}, \label{sf_equation}
\end{equation}
where $N(\tau)$ is the number of pairs separated by a time difference $\tau$, and $m_i - m_j$ the magnitude difference for each pair $N(\tau)$. We discretised Eq. (\ref{sf_equation}) using logarithmically spaced bins. Each bin corresponds to a range of time differences $(\tau_1, \tau_2)$, and is represented by its midpoint $\tau$. We selected all pairs of observations whose time separation $\Delta t = t_i - t_j$ lies within this range to compute SF$(\tau)$ at each bin. Bins that do not contain any pairs are excluded from the analysis. To further adapt this binning strategy to unevenly sampled light curves, we introduced a tolerance factor $\epsilon$, which ensures that each bin contains enough pairs for a reliable calculation \citep{Roelens2017}. If a populated bin initially has too few pairs, its size is increased symmetrically until the minimum requirement of at least three pairs is met. With GDR3 data, we could compute SFs for time differences in the range 0.07--980 d (see Fig.~\ref{fig_sf_all}). The number of bins used was based on the number of epochs available for each light curve, with 1.5 times the number of epochs used for the number of bins. 

For aperiodic variability, the SF typically shows three main regimes. A noise-dominated regime where the SF remains roughly constant around the intrinsic noise level in the data. A variability regime where the SF value increases until it peaks at the turnover timescale $t_0$. And a plateau where, beyond $t_0$, the SF flattens as no new variability occurs. To represent these three regimes, we fitted the SFs with a function of the form

\begin{equation}
    y = C_0 + C_1(1 - \exp{(-\tau/t_0)}), \label{SF_fit_equation}
\end{equation}

\noindent where $C_0$, $C_1$, and $t_0$ are the fitting parameters and $\tau$ is the time difference. In this model, $C_0$ is related to intrinsic variability due to the uncertainty of the measurements and $C_1$ is associated with the real signal variability amplitude. The final fitting parameter $t_0$ is the turnover timescale, which marks the transition between the variability-dominated regime and the plateau and can be interpreted as the variability timescale in which we are interested.

We fit Eq. (\ref{SF_fit_equation}) to the SF using the \texttt{MIGRAD} routine \citep{James:1975dr} implemented in Python as part of the \texttt{iminuit} library \citep{iminuit}. The initial guesses and limit values for the model parameters were based on inspection of the distribution of all SF shown in Fig.~\ref{fig_sf_all} and were chosen as follows: $C_0 = 0.02$ , $C_1 = 0.08$, $t_0 = 15$~d for the initial guess. 
While $C_0$ is in the range $(10^{-5}, 0.07)$ mag$^2$, $C_1$ in $(10^{-3}, 5.5)$ mag$^2$ and $t_0$ in $(0.07, 980)$~d. Examples of SF model fit results can be seen in Fig. \ref{fig_sf}. The imposition of model parameter limits resulted in a relative excess of fit results with edge values. Hence, we filtered these results to remove spurious models within 5\% of the edge values for $C_0$, $C_1$, and for the lower limit of $t_0$. A total of 5\,354 SF fits were discarded at this step. Visual inspection of discarded models showed that these were typically sources with complex multi-mode variability, including different timescales and that could not be represented by Equation (\ref{SF_fit_equation}), or sources with poorly sampled SF with a long time range without data (blind time range). An excess of upper limit values of $t_0$ also existed, and we verified that these were typically related to light curves exhibiting a long-term trend (e.g. a gradual change in brightness) with a timescale longer than the time span of GDR3 observations, where the SF continues to increase without ever reaching the plateau, in which case, we stored the timescale as evidence for long-term variability but did not use them to evaluate the Q index. In addition to the issue of edge values, 12 best-fit models were also discarded because they had limited $\tau$ sampling around the derived timescale. This issue is further discussed in the App.~\ref{app_SF_tau}, and it is introduced by GDR3 sparse time-series sampling, which for some sources results in a blind time range in the SF. For affected sources (see Fig.~\ref{fig_sf}-bottom), no data points exist in the range $\tau\sim 0.3-28$ days. Beyond this range, the SF show a flat distribution - hence, it is impossible for models to correctly identify the transition from the variability regime into the plateau.

The fits of the SF model in this section allowed us to estimate the aperiodic variability timescale for $4\,118$ sources. For $1\,245$ sources, the SF-fit resulted in upper-edge values or timescales that were longer than half of the total time span of the light
curve, with these stored as evidence of long trend, but were not used to evaluate the Q index. For $5\,363$, we were unable to derive a time scale of variability using the GLS or SF methods.

\subsection{Residual spurious variability}\label{sec:residual_spur_var}

 We further discarded 1\,251 sources as their measured timescale coincided with typical timescales caused by the spurious variability introduced by the Gaia scan-angle effects (Sect.~\ref{sec:spurious_var}, see also App.~\ref{spurious_residual}). After this final cleaning step, we proceeded with the analysis of 28\,179 YSO light curves, with 9\,824 of these in the primary sample and 18\,355 in the secondary sample. 

\subsection{Q and M implementation and morphology classes} \label{subsection_morpho_class}

Once a characteristic timescale was derived, light curves were folded to their corresponding phase, and a waveform was derived using a Savitzky-Golay filter with a window size of 25\% of the phase and a second-order polynomial. This set of parameters allowed us to obtain waveforms representative of the ongoing variability while avoiding overfitting. The estimated waveform was then subtracted from the data to obtain the residual curve, followed by estimating the Q indices with Eq. (\ref{Q_eq}).

Following \citet{Cody_14} scheme, the Q and M indices can be used to gauge light curve morphologies and sort variables into seven standard classes:

\begin{itemize}
    \item EB:  $Q<0.11$ and $M>0.25$.
    \item P: $Q<0.11$ and $M\leq0.25$.
    \item QPS: $0.11\leq Q\leq0.61$ and $|M|\leq0.25$.
    \item QPD: $0.11\leq Q\leq0.61$ and $M>0.25$.
    \item APD: $Q>0.61$ and $M>0.25$.
    \item B: $Q>0.11$ and $M<-0.25$.
    \item S: $Q>0.61$ and $|M|\leq0.25$.
\end{itemize}

Additional classes are sometimes included in the literature, such as multi-periodic, non-variable, and L. These classes are not based directly on the Q and M indices, but are defined based on the failure to apply Q-M method. The identification of multi-periodic variability in previous studies relies on the high cadence of their observations and requires additional steps, including visual inspection of the light curves. We did not reproduce those steps as part of this study. As our sample is focused on light curves with enhanced variability, the non-variable YSOs are also excluded. L variables were identified here as sources with inferred timescales longer than half of GDR3 time span.

\section{Results}\label{section_results}

\subsection{Asymmetry and periodicity indices}

\begin{figure}[htb]
   \centering
      \includegraphics[width=1\linewidth]{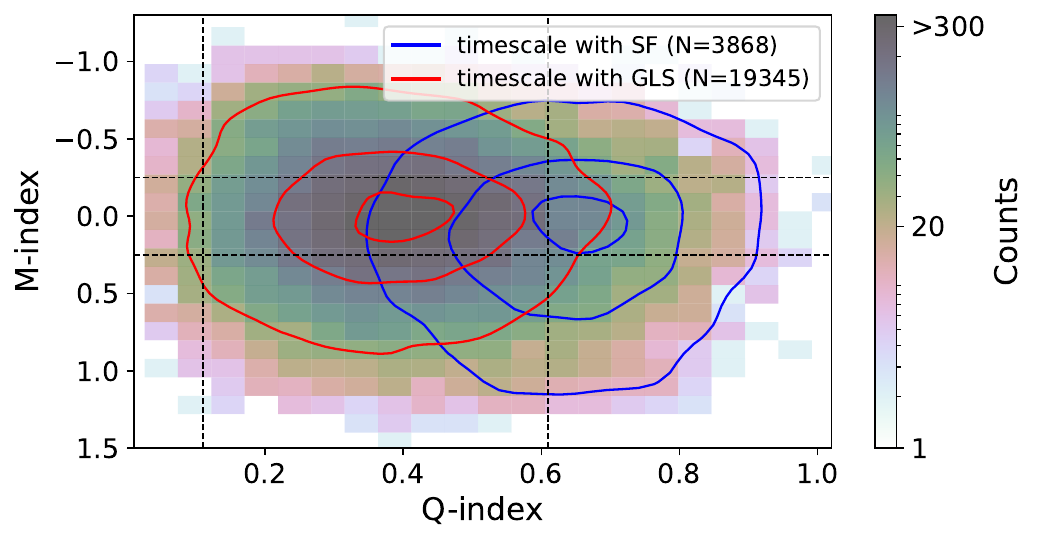}
      \caption{Distribution of Q and M indices the GDR3 YSO sample. Red a blue lines contour the distribution of sources with GLS periodic timescale, and SF aperiodic timescale respectively (Sect.~\ref{sec:timescale}). The dashed lines show the adopted thresholds Q-M classification. The three contour curves contain 10, 50, and 90\% of the data points, respectively.}
         \label{fig_QM}
   \end{figure}

\begin{figure*}[htb]
    \centering
    \includegraphics[width=\textwidth]{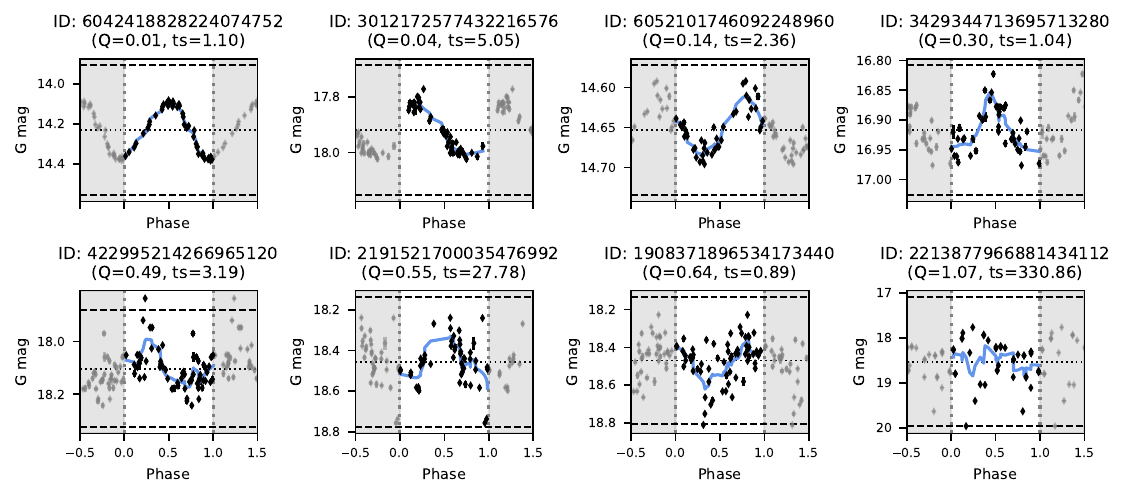}
    \caption{Phase-folded GDR3 light curves are shown with Q values increasing from left to right. Black diamonds indicate the phase-folded epoch photometry, while the blue line represents the inferred waveform. The grey region marks the extended phase for clarity. Grey horizontal lines denote the mean (dotted) and the mean $\pm 3\times$ std (dashed).}
    \label{fig:Q_evolution}
\end{figure*}

\begin{figure*}[htb]
    \centering
    \includegraphics[width=\textwidth]{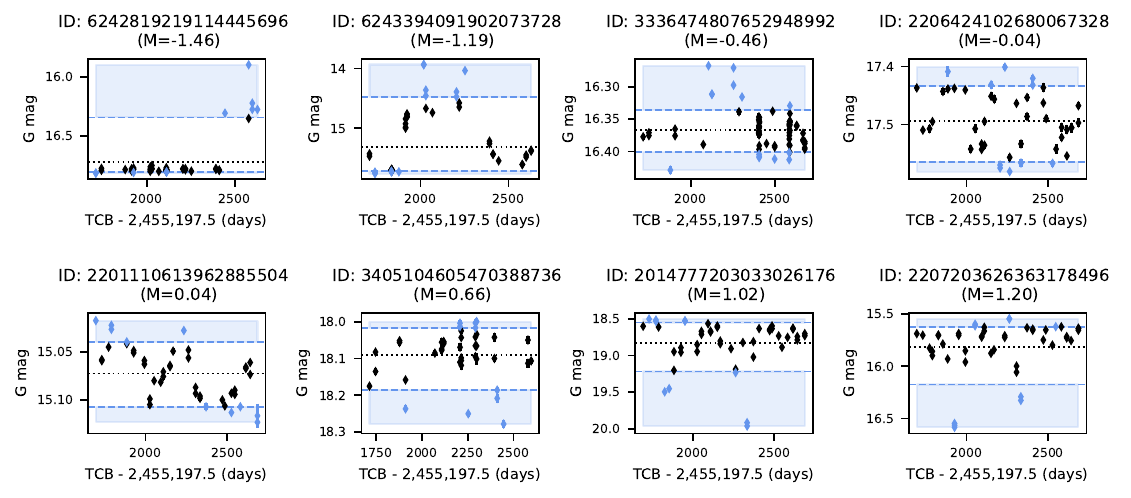}
    \caption{Examples of GDR3 light curves with increasing M values from left to right. The blue diamonds are the data points of the top and bottom deciles of the curve; the grey dotted line shows the mean; the blue dashed lines show the top and bottom deciles. The time axis reflects the observations'  Barycentric Coordinate Time (TCB).}
    \label{fig:M_evolution}
\end{figure*}

We successfully derived Q indices for 23\,213 sources (8\,385 in the primary sample) and M indices for 28\,179 sources (9\,824; primary). Our successful adaptation of the Q-M method for use with GDR3 light curves is illustrated in Fig.~\ref{fig:Q_evolution} and Fig. \ref{fig:M_evolution}. In Fig.~\ref{fig:Q_evolution}, light curves at low Q values are very well represented by the estimated periodic waveform. At intermediate Q values, quasi-periodic waveforms are still found, but with increasing phase dispersion. At $Q\approx 1$, light curves are dominated by stochastic variability, timescales inferred can no longer be interpreted as a periodic timescales, with data points are stochastically distributed around the average. In Fig. \ref{fig:M_evolution}, light curves with extreme asymmetries have M values farthest away from 0, with negative M values corresponding to variables with brightening events and positive M values corresponding to dimming events, while more symmetric light curves are closest to $M\approx0$.

\begin{figure}[htb]
   \centering
      \includegraphics[width=1\linewidth]{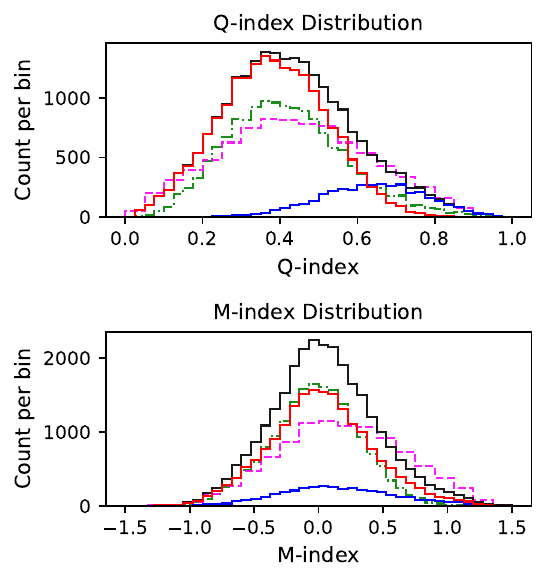}
      \caption{Histograms for the distributions of the Q (top) and M (bottom) indices. L variables are included in the M distribution, but excluded from the Q distribution. The full sample is shown in black, sources with periodic GLS timescales in red, and those with aperiodic SF timescales in blue. The primary and secondary samples appear as magenta dashed and green dash-dotted lines, respectively.}
         \label{fig_QM_hist1d}
   \end{figure}

Fig. \ref{fig_QM} shows the distribution of Q versus M for our sample. This general distribution peaks at $Q\sim0.43$ (quasi-periodic) and $M\sim0.05$ (symmetric). The individual distributions of the Q and M indices are shown in Fig.~\ref{fig_QM_hist1d}, where we separate our sample by timescale method. As expected from the nature of the methodology for measuring variability timescales, the distribution of Q for sources with a periodic timescale derived with the GLS (in red) is mainly on the periodic and quasi-periodic regions of the diagram, peaking at $Q\sim0.38$, with the Q index distribution for sources with aperiodic timescales derived from SFs peaking at $Q\sim0.64$. For the M index, sources with periodic timescales peak at $\sim$0.02, and at $\sim$0.15 for aperiodic timescales, with the aperiodic distribution showing a tail towards dipping M index values. Fig.~\ref{fig_QM_hist1d} also compares the distribution of Q-M indices for our primary and secondary samples, where the differences for these two are also supported by Kolmogorov–Smirnov (KS) tests. Quasi-periodic variability is the most frequent mode in our sample, accounting for more than 69\% of sources. The secondary sample peaks at $Q\sim0.38$ and  $M\sim0.02$, being dominated by QPS sources. The over-representation of low-amplitude QPS sources in the general occurrence of variability classes motivated discussing results as a function of higher-amplitude (primary) vs lower-amplitude (secondary) variable samples. The primary sample of variables peaks at $Q\sim0.44$ and $M\sim0.18$, with a larger proportion of dipping and aperiodic variables in comparison to the secondary sample.  

The classification scheme outlined in Sect.~\ref{subsection_morpho_class} was applied to sort light curves into seven morphological classes with an eighth class attributed to account for L variables, which had characteristic timescales longer than half the total time span of the GDR3 observations (typically $\sim475$ days). The count and frequency of sources in each class are summarised in Table~\ref{tab:gdr3_allsky}. As already suggested by the Q-M distributions, dipper variability is the most common among the primary sample, with $\sim$38\% of these sources showing APD or QPD variability. We identified 4\,966 L variables, 3\,721 with periodic timescales derived with the GLS and 1\,275 aperiodic timescales derived with the SF. We visually inspected a large fraction of these light curves and confirmed that a long-term trend was indeed present. Their timescales are at the limit of GDR3 resolvable timescales and should be interpreted as a bottom limit for the real timescale, and hence we did not employ them to calculate the Q index, although we discuss them as part of our results.

\subsection{Amplitude and timescales}

\begin{figure}[htb]
   \centering
      \includegraphics[width=1\linewidth]{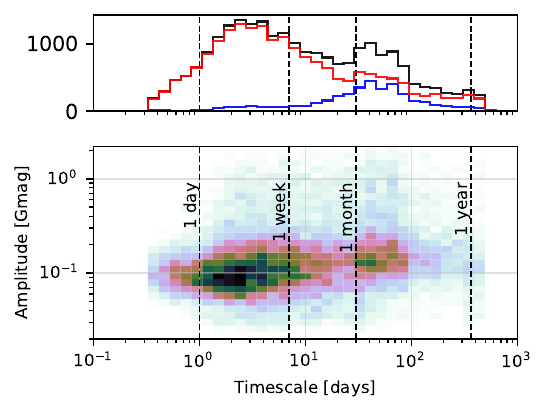}
      \caption{{Top:} Timescale distribution for the full sample (black), 19\,345 sources with GLS timescale (red), and 3\,868 sources with an SF timescale (blue) {Bottom:} Amplitude vs. timescales for the full sample of 23\,213 sources with a derived timescale. L variables are excluded.}
         \label{fig_ts_q}
   \end{figure}

We further complemented our results analysis with peak-to-peak (ptp) amplitudes and timescales measurements. We used the min-max amplitude as a measure of the ptp amplitude rather than the 5-95 percentile amplitude more commonly employed in the literature, as we found the latter to underestimate ptp amplitudes when applied to GDR3 sparse sampling. Throughout the discussion of results, we refer to timescales in the range $0.5-16$ d as ``rotation timescales''. Timescales larger than $20$ d and less than about half of the duration ($\sim 475$ days) of the observations are termed ``intermediate timescales''. Timescales longer than half of the duration of the light curve are referred to as ``long-trend timescales''. 
Fig.~\ref{fig_ts_q} shows the distribution of amplitudes as a function of timescales for our sample, revealing a bimodal distribution of timescales with a population of sources with rotation-timescales ($62\%$) and a second population of variables with intermediate timescales ($\sim24\%$ of the sample).

\begin{table*}[bht]
\caption{Counts and frequencies of sources for each variability morphologic class}
\centering
\begin{tabular}{r|r|ccc|cc|rrrr}

\hline
 & & \multicolumn{3}{|c|}{Frequency\tablefootmark{a}}  & & & \multicolumn{4}{|c}{Spectral Type\tablefootmark{b}}\\
{Class} & {count} & {thick disc} & {general disc} & {CTTS}  & ts-ratio  & ptp & M & K & G & AB-F\\
\hline
\multicolumn{3}{l}{Full sample}  & & & & & \\
P	& 286  & 14.3 (2.1) & 50.0 (3.0) &	13.7 (2.8) & 13.4 &	0.19 & 77.3	(2.5) & 12.9 (2.0)	& 1.7 (0.8) & 8.0 (1.6)\\
QPS	& 9514 & 20.0 (0.4) & 57.6 (0.5) & 14.8 (0.6) & 2.6 &	0.15 & 71.2 (0.5) & 15.3 (0.4) & 4.0 (0.2) & 9.5 (0.4)\\ 
QPD	& 5439 & 39.1 (0.7) & 67.3 (0.6) & 24.7 (1.1) & 1.8 &	0.35 & 65.4 (0.6) & 18.6 (0.5)& 5.4 (0.3) & 10.6 (0.4)\\ 
APD	& 1103 & 56.6 (1.5) & 79.9 (1.2) & 39.7 (3.7) & 0.4 &	0.53 & 66.2	(1.4) & 19.7 (1.2) & 5.7 (0.7) & 8.4 (0.8)\\
B	& 5252 & 21.0 (0.6) & 56.7 (0.7) & 22.0 (1.1) & 2.0 &	0.16 & 59.8	(0.7) & 24.3 (0.6) & 3.2 (0.2) & 12.7 (0.5) \\
S	& 1490 & 27.9 (1.2) & 65.9 (1.2) & 23.6 (2.6) & 0.8 &	0.19 & 71.4	(1.2) & 17.8 (1.0) & 3.2 (0.5) & 7.6 (0.7)\\
EB	& 129 &	15.5  (3.2) & 56.6 (4.4) & 11.4 (3.6) & 23.6 &	0.20 & 85.3	(3.1) & 9.3 (2.6) & 0.8 (0.8) &	4.6 (1.9) \\
L	& 4966 & 21.1 (0.6) & 48.5 (0.7) & 11.0 (1.1) & - & 	0.16 & 44.3 (0.7) & 22.9 (0.6) & 7.7 (0.4) & 25.1 (0.6)\\
\hline

\multicolumn{3}{l}{Primary sample}   & & & & & \\
P   &  199 &  17.1 (2.7) & 46.2	(3.5) & 14.6 (3.1) &  26.6&   0.22 & 81.4 (2.8) & 8.6 (2.0) & 2.0 (1.0) & 8.0 (1.9) \\
QPS &  2362 &   41.9 (1.0) & 65.5 (1.0) & 36.6 (1.6) &   1.4&   0.31 & 60.5 (1.0) & 22.2 (0.9) & 7.1 (0.5) & 10.2 (0.6) \\
QPD &  2886 &   61.7 (0.9)& 81.4 (0.7) & 40.6 (1.6) &   1.3&   0.57 & 65.5 (0.9) & 20.5 (0.8) & 6.1 (0.4) & 7.9 (0.5)\\
APD &  822  &  68.0 (1.6) & 85.2 (1.2) & 48.2 (4.2) &  0.4 &  0.67 & 63.4 (1.7) & 21.9 (1.5) & 7.0 (0.9) & 7.7 (0.9)\\
B   &  1505 &   44.8(1.3) & 69.4 (1.2) &45.6 (2.2) &   1.3&   0.31 & 68.4 (1.2) & 18.2 (1.0)& 4.3 (0.5)	& 9.1 (0.7)\\ 
S   &  601  &  42.8 (2.0) & 70.9 (1.9) & 49.0 (4.9) &  0.4 &  0.30 & 62.1 (2.0) & 23.3 (1.7) & 4.8 (0.9) & 9.8 (1.2)\\
EB  &  95   & 15.8  (3.7)  & 51.6 (5.1) & 14.1 (4.3) & 91.0 &  0.24 & 84.2 (3.7)& 8.4 (2.8)& 1.1 (1.1) & 6.3 (2.5) \\
L   &  1354 &   40.0(1.3) & 63.1 (1.3) & 33.6 (3.2) &   &   0.37 & 48.6	(1.4)& 26.3 (1.2) & 8.2 (0.7) & 16.9 (1.0)\\

\hline
\multicolumn{3}{l}{Secondary sample}  & & & & & \\
P   &   87 &   8.0   (2.9) & 58.6 (5.3) & 8.7 (5.9) &   5.8  &   0.10 & 67.8 (5.0) & 23.0 (4.5) & 1.1 (1.1) & 8.0 (2.9)\\
QPS & 7152 &   12.8  (0.4) & 55.0 (0.6) & 6.1 (0.5)&    3.3 &    0.09 & 74.7 (0.5)& 13.1 (0.4) & 3.0 (0.2) & 9.2 (0.3) \\
QPD & 2553 &   13.4  (0.7) & 51.5 (1.0) & 4.5 (0.8) &    2.5 &    0.09 & 65.2 (0.9)& 16.5 (0.7) & 4.6 (0.4)& 13.7 (0.7)\\
APD &  281 &   23.1  (2.5) & 64.4 (2.9) & 3.0 (3.0) &    0.7 &    0.13 & 74.4 (2.6) & 13.5 (2.0) & 1.8 (0.8) & 10.3 (1.8)\\
B   & 3747 &   11.4  (0.5) & 51.6 (0.8) & 7.0 (0.9) &    2.4 &    0.10 & 56.4 (0.8) & 26.8 (0.7) & 2.6 (0.3) & 14.2 (0.6)\\
S   &  889 &   17.8  (1.3) & 62.5 (1.6) & 7.4 (2.0) &    1.1 &    0.12 & 77.8  (1.4)& 14.0 (1.2) & 2.1 (0.5) & 6.1 (0.8) \\
EB  &   34 &   14.7  (6.1) & 70.6 (7.8) & - &    6.8 &    0.10 & 88.2 (5.5)	& 11.8 (5.5) & - & - \\
L   & 3612 &   14.0  (0.6) & 43.0 (0.8) & 2.5 (0.7) &    -   &    0.09 & 42.6 (0.8) & 21.6 (0.7)& 7.6 (0.4) & 28.2 (0.7)\\

\end{tabular}
\tablefoot{Morphologic classes defined as in Sect.~\ref{subsection_morpho_class}. The primary and secondary samples are detailed in Sect.~\ref{sec:selection_high_var}. Thick-disc frequency indicates the frequency of stars with an envelope or a thick disc, as indicated by the $\alpha_\mathrm{IR}\geq -1.6$. General disc frequency refers to the frequency of sources with $\alpha_\mathrm{IR}\geq -2.5$ or with EW(H$\alpha$) indicative of ongoing accretion. CTTS frequency indicates the frequency of sources identified as CTTS in Sect.~\ref{sec:Ha} in comparison with the total number of sources with $H\alpha$ information. ts-ratio indicates the ratio of sources with rotation against intermediate timescales. ptp indicates the average peak-to-peak amplitude per class. The primary and secondary sample are detailed in Sect.~\ref{sec:selection_high_var}. \tablefoottext{a}{ Values in parentheses are standard error estimates for the disc frequencies.} \tablefoottext{b}{Frequency of spectral types as a function of variability class}}
\label{tab:gdr3_allsky}
\end{table*}

As detailed in Sect.~\ref{sf_ts} and App.~\ref{app_SF_tau}, a combination of factors such as the sparse sampling of GDR3 and the SF discretisation process makes the SF method inefficient for detecting aperiodic timescales in the range $\sim 0.3-28$ d. Hence, if the GLS method was inefficient at detecting long timescales, this bi-modality could be explained by the detection limits introduced by the combined use of these two methods. Conversely,  Fig.~\ref{fig_ts_q} shows that the distribution of timescales measured with the GLS method is itself bi-modal, with 28\% of GLS-derived values at intermediate-timescales. Additionally, albeit much reduced in numbers, the SF timescale distribution accounts for $\sim6\%$ of rotation timescales. Another possible explanation would be that the intermediate-timescale population is the result of the spurious variability discussed in Sect.\ref{sec:spurious_var}. However, we have addressed and mitigated the occurrence of residual spurious variability in our sample in App.~\ref{spurious_residual} and Sect.~\ref{sec:residual_spur_var}, where we discarded sources with timescales in the ranges typically susceptible to spurious variability. Finally, when comparing our results with previous studies using ZTF light curves (Sect.~\ref{ground-based}), we found that those previous studies also find a hint to similar bi-modality in timescales, with $\sim74\%$ vs $\sim9\%$ incidence of rotation and intermediate timescales. We are therefore confident that the observed bi-modality is real. The two populations also show distinctive behaviour in ptp amplitudes, with rotation timescales showing average amplitudes of 0.18 mag (standard deviation 0.26) and intermediate timescales at 0.29 mag (0.38), and these differences are also supported by a KS test. 

\subsection{Asymmetry and periodicity metrics in the context of the observed YSO properties}\label{sec:QMmetrics}

In this section, we examine our results in the context of quantities related to the physical characteristics of YSOs.

\subsubsection{Accreting versus non-accreting sources}\label{sec:Ha}

\begin{figure}[htb]
   \centering
   \includegraphics[width=.9\linewidth]{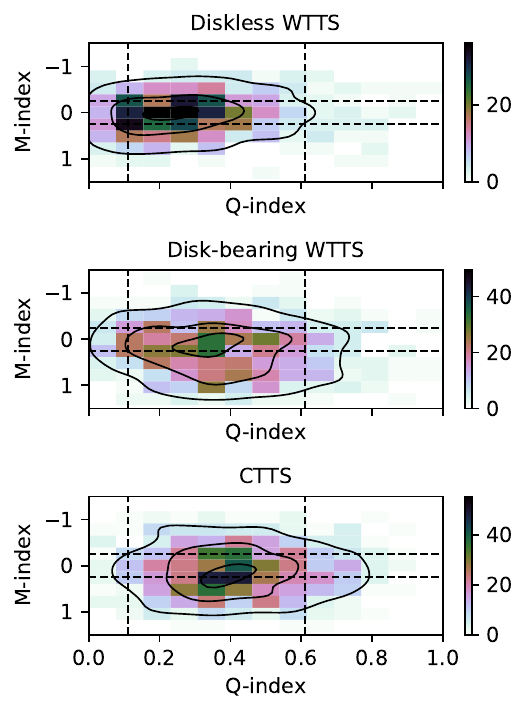}
    \caption{Q-M distributions for subsamples of sources with no observational indication of ongoing accretion (WTTS) and absence of a disc as identified from their IR SED (top), non-accreting sources (WTTS) whose IR SED indicates a disc (middle), and sources identified as actively creating (CTTS, bottom).} \label{fig_qm_ha}
\end{figure}

In App.~\ref{app:accretion}, we classified sources as actively accreting Classical T Tauri Stars (CTTS) or non-accreating weak-lined TTS (WTTS) based on their H$\alpha$ emission. In total, we could assign CTTS/WTTS labels to 13\,802 GDR3 YSOs, with 7\,738 in the variable sample. Among 2\,829 sources in the primary sample with both Q and M derived, 1\,107 were CTTSs and 1\,722 were WTTSs. Their Q-M distribution is shown in Fig.~\ref{fig_qm_ha}, where we further distinguish between disc-bearing and discless WTTS based on their $\alpha_\mathrm{IR}$ index (Sect.~\ref{sec:alpha_ir}). Sources with H$\alpha$ information have a lower incidence of aperiodic variability ($\sim 11\%$) compared to the full variable sample ($\sim20\%$).

The Q-M distribution for discless WTTS peaks at $Q\sim0.29$ and $M\sim0.04$, with Q values shifted toward the intersection between strictly and quasi-periodic sources, and a narrower distribution of M values concentrated toward symmetric M indices. The subsample of actively accreting sources (CTTS) is mainly quasi-periodic, with their distribution peaking at $Q\sim0.42$ and $M\sim0.13$. Compared to the discless WTTS distribution, CTTSs are shifted towards larger Q values, with a much wider distribution of M values including more significant fractions of dippers and bursters. For disc-bearing WTTSs, $Q\sim0.38$ and $M\sim0.24$, with the distribution presenting intermediary properties between the former two groups. Disc-bearing WTTSs show a smaller relative fraction of P sources compared to discless WTTS as well as an excess of QPDs with higher M values than CTTSs.

To further support the discussion of the results, Table~\ref{tab:gdr3_allsky} includes the CTTS frequency defined as the percentage of CTTS sources among all sources with H$\alpha$ measurements, and interpreted here as a measure of the fraction of sources actively accreting within a sample. We note that H$\alpha$ measurements are limited to optically visible sources. Consequently, this proxy will inevitably miss deeply embedded sources. Furthermore, we also note that the accretion process and the resulting H$\alpha$-emission are also highly variable \citep[e.g.][]{Sousa2016A&A...586A..47S}. Hence, although strong H$\alpha$ emission indicates ongoing accretion, the opposite is not always true, as sources may still be sporadically accreting over longer timescales although not in the specific season observed. CTTS fraction must thus be interpreted as a bottom limit of the true fraction of actively accreting sources.

\subsubsection{Circumstellar discs}\label{sec:alpha_ir}

\begin{figure}[htb]
   \includegraphics[width=.9\linewidth]{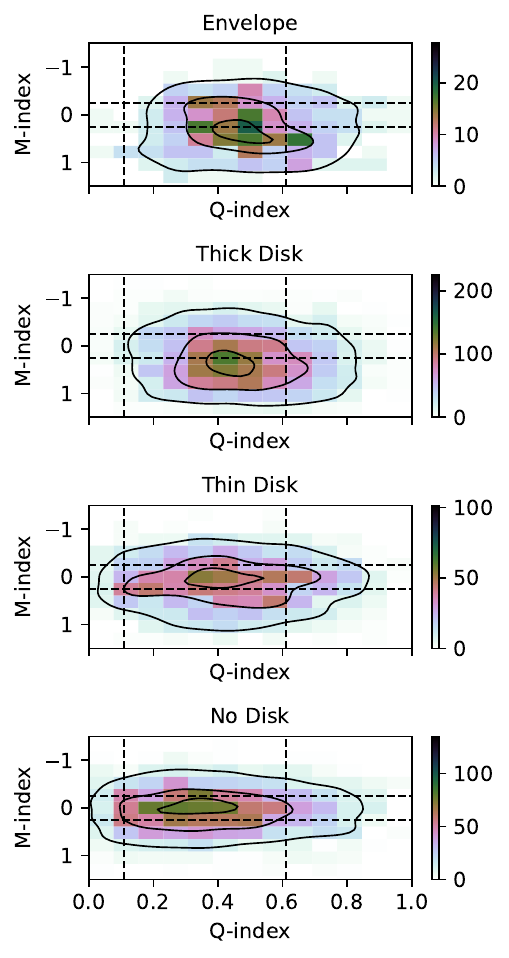}
    \caption{Q-M distributions for subsamples of sources with $\alpha_\mathrm{IR}$ index (see Sect.~\ref{sec:alpha_ir}) indicative of contribution of an envelope to their SED (top), a thick-disc (middle-top), a thin disc (middle-bottom), or discless sources (bottom). Colour maps reflect the source count per bin.}
    \label{fig_qm_alpha}
\end{figure}

To gauge the influence of circumstellar discs on our results, we examined the $\alpha_\mathrm{IR}$ indices for our sources. The $\alpha_\mathrm{IR}$ index is a measure of the slope of the IR spectral energy distribution (SED) and quantifies the contribution to the SED of infrared radiation re-emitted by material around YSOs. { In App.~\ref{app:alpha_ir},} we were able to derive $\alpha_\mathrm{IR}$ for 27\,653 sources (98\% of our final sample). { To facilitate our discussion, we used $\alpha_\mathrm{IR}$ to classify sources according to the dominant contributor to their SED: 
712 sources in the primary sample and 697 in the secondary had a significant envelope contribution; 
4\,296 (primary) and 2\,131 (secondary) sources had a thick disc; 2\,021 and 6\,722 sources had a thin discs; and 2\,795 and 8\,836 sources were likely discless.} Fig. \ref{fig_qm_alpha} shows the Q-M distributions for each of these samples. Table~\ref{tab:gdr3_allsky} includes the frequency of sources with $\alpha_\mathrm{IR}$-based evidence for the existence of an inner disc, which we define as the total number of sources classified as envelope or thick disc, divided by the total number of sources with $\alpha_\mathrm{IR}$ index. 

KS tests between envelope and thick disc samples suggest that these samples may share the same mother Q-M distribution. However, when examined in terms of the ratio between rotation and intermediate timescales, we find a 0.7 ratio for the envelope sample compared to a 1.0 ratio for the thick disc sample. Generally, Fig.~\ref{fig_qm_alpha} suggests a gradual transition towards a distribution of M indices that favours symmetry and a shift toward enhanced periodicity for more evolved YSO populations. 

Several of the variability mechanisms in YSOs stem from the star-disc interaction process, where the young star magnetically interacts with its inner disc. Hence a relevant quantity included in Table~\ref{tab:gdr3_allsky} is the percentage of sources with $\alpha_\mathrm{IR}$ index indicative of the presence of a inner disc, which we define as $\alpha_{IR}\geq -1.6$ and includes both sources with a thick disc or an envelope. { We also report the fraction of sources with evidence for circumstellar material (namely general disc-fraction), which we define as the fraction of sources  $\alpha_\mathrm{IR}\geq-2.5$ or that had EW(H$\alpha$) indicative of accretion (Sect.~\ref{sec:Ha}.)}. Our $\alpha_\mathrm{IR}$ are derived based on heterogeneous data available from previous publications and thus may be biased toward a lack of mid- and far-infrared data, which are only available all-sky within the context of \emph{WISE} \citep{Wright2010_WISE}, with $\sim31\%$ of our compiled SEDs including only data below $5\mu m$. We therefore stress that reported disc frequencies represent a lower bound on the actual frequencies. For example, the thick disc fraction for the full sample goes from 25.5\% to 37.3\% when considering only sources with SEDs populated beyond $\lambda\gtrsim5\mu m$. For a discussion of how using SEDs without data beyond $\lambda\gtrsim5\mu m$ affects studies based on $\alpha_\mathrm{IR}$, see \citet[][app. C]{roquette2024}, who report a 25.2\% impurity among thin disc sources identified from such undersampled SEDs.

\subsubsection{Spectral types}

Fig.~\ref{fig:SpT_QM_freq} shows the distribution of spectral types for our sample, estimated in App.~\ref{app:SpT}, illustrating that $\sim 83\%$ of our sample is composed of K and M sources, with a smaller incidence ($\sim13\%$) of intermediate mass YSOs. In Table ~\ref{tab:gdr3_allsky} we present the incidence of spectral types in each variability class, and in Fig.~\ref{fig:SpT_QM_freq}, the incidence of variability classes for each spectral type. The latter, in particular, illustrates two trends as a function of spectral type. Earlier spectral types are more frequently classified as L variables, whereas QPS sources occur more often at later spectral types. We also note that B, QPD and APD also vary with spectral type, although variations are much smaller and show a less clear trend. In our discussion of the results, we group A, B and F type objects as intermediate-mass or early-type YSOs. G , K, and M type objects are treated as late-type sources, with K and M type sources distinguished when relevant.

\begin{figure}
    \centering
    \includegraphics[width=0.95\linewidth]{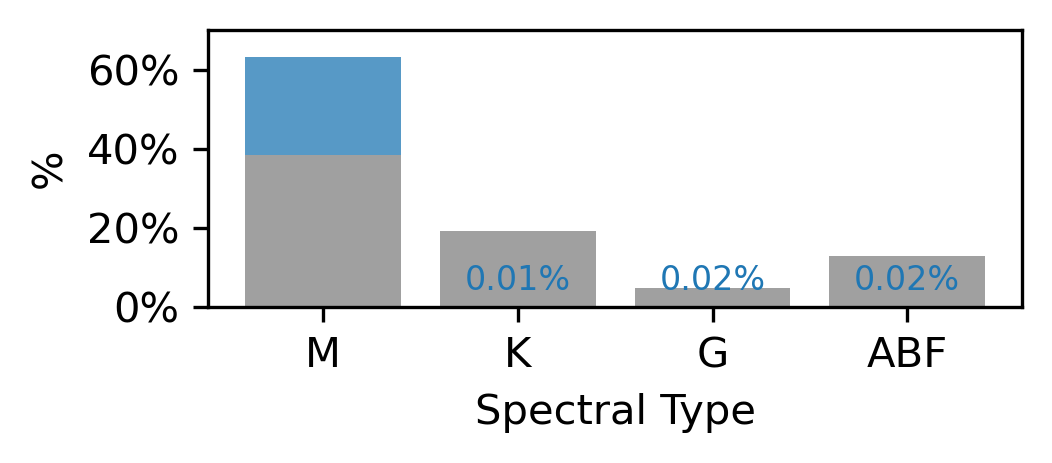}
    \caption{Percentage of sources in the YSO variable sample as a function of their spectral types. Spectral types derived in App.~\ref{app:SpT} from their GSP-Phot effective temperatures are shown in grey, and in blue if derived from de-redden $G_{BP}-G_{RP}$ colours. Bars too small for visualisation have their fractional contribution written in the plot.}
    \label{fig:SpT_freq}
\end{figure}

\begin{figure}
    \centering
    \includegraphics[width=0.9\linewidth]{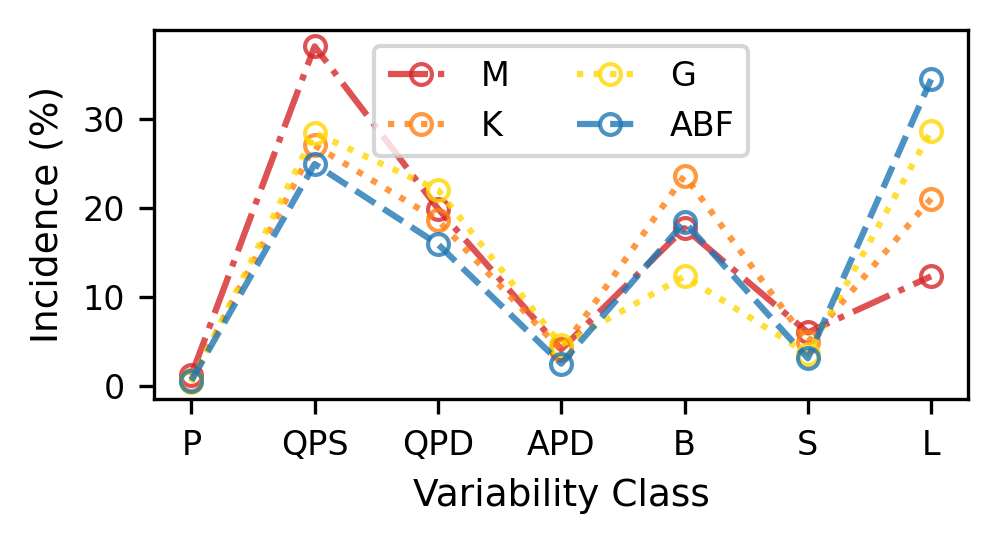}
    \caption{Incidence of variability classes as a function of spectral type.}
    \label{fig:SpT_QM_freq}
\end{figure}

\section{Discussion}\label{section_discussion}

The effectiveness of the Q-M metrics for characterising YSO variability has already been demonstrated in the literature with successful applications to both high-cadence space-based and longer-term ground-based observations. In this study, we extend prior work by showing that Q–M metrics can be applied to Gaia’s sparse yet highly precise long-term data. Having established this, we next consider how to interpret our findings within the Q–M morphological classification scheme.

\subsection{Incidence of morphological classes across surveys}\label{sec:class_frac_discussion}

\begin{figure*}[h]
    \centering
    \begin{tabular}{ccc} 
        \includegraphics[width=0.32\textwidth]{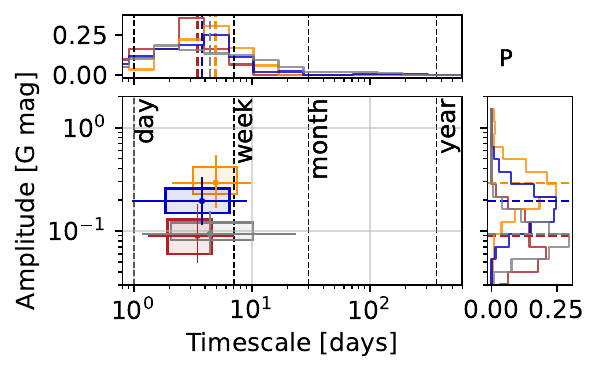}&
        \includegraphics[width=0.32\textwidth]{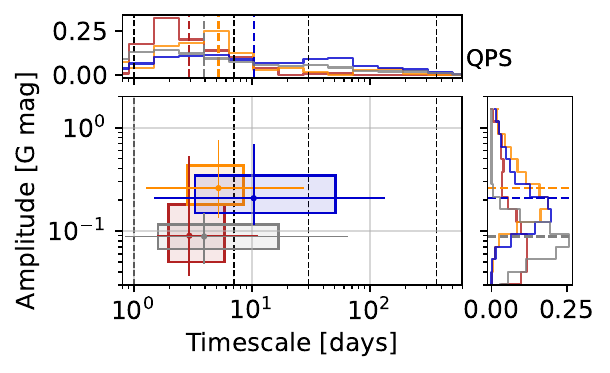}&
        \includegraphics[width=0.32\textwidth]{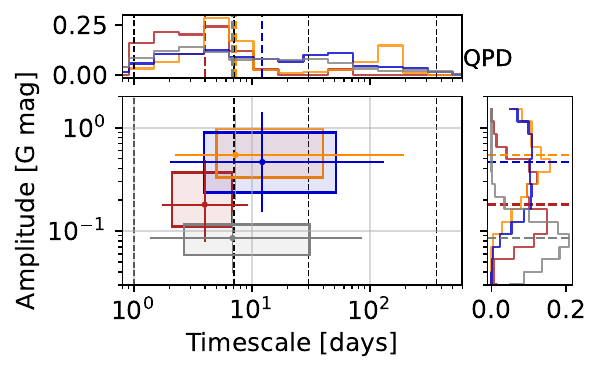}\\ 
        
        \includegraphics[width=0.32\textwidth]{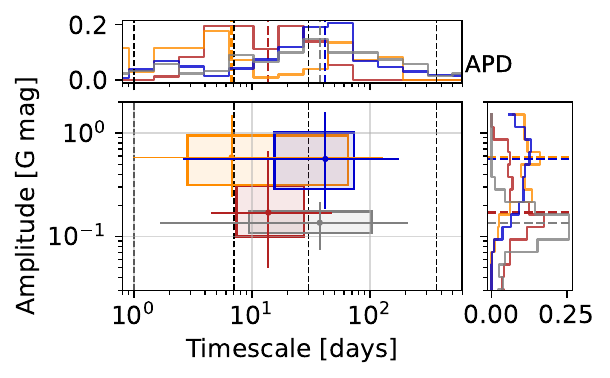}&
        \includegraphics[width=0.32\textwidth]{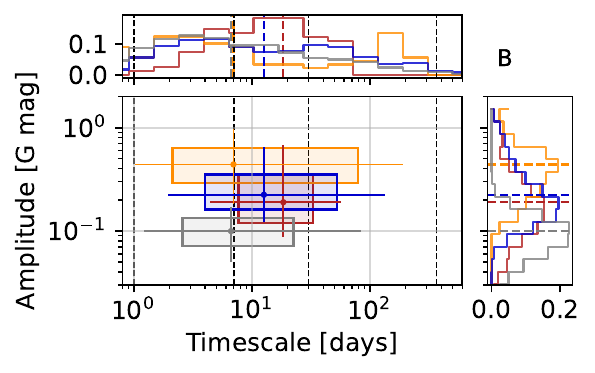}&
        \includegraphics[width=0.32\textwidth]{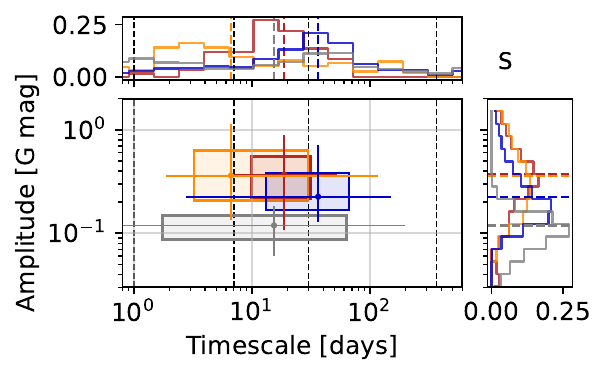} \\
    \end{tabular}
    \caption{Box plots of the timescale–amplitude distributions are shown for six morphology classes defined by the Q–M indices, along with normalised histograms for each parameter. The primary sample is plotted in blue and the secondary sample in grey. High-cadence space-based sources (CoRoT, K2) and ground-based ZTF results are shown in red and orange. }
    \label{fig:Q-M_boxplots_comp}
\end{figure*}

Table~\ref{tab:comp_gdr3_previous_studies} compares the population-level frequencies of variability morphologies reported in different works. Morphological classes from the literature are taken as reported, without review of their Q-M thresholds. Some previous studies included extra classes for multi-periodic and non-variables, which we did not consider here and we estimated their class incidences excluding these two categories. We thus stress that these frequencies reflect the incidence of morphological classes among samples of variable stars. Fig.~\ref{fig:Q-M_boxplots_comp} presents amplitudes versus timescales distributions for each morphological class. This includes our primary and secondary samples, as well as the main samples from the literature combined by type of survey, including high-cadence surveys \citep[CoRoT and K2;][]{Cody_14, Cody_2018, Venuti_2021, Cody_2022} and ground-based results \citep[ZTF;][]{Hillenbrand2022, Wang2023, JiangHillenbrand2024}.  Detailed discussion of each morphological class are later presented in Sect.~\ref{sec:QMclasses}.

\begin{table}[htb]
\caption{Incidence (in per cent) of the variability classes for GDR3 light curves in comparison with previous studies.}
\centering
\begin{tabular}{l|rrr|rr}
\hline
&  \multicolumn{3}{c}{This Work} & \multicolumn{2}{c}{Previous Studies\tablefootmark{a}} \\
{} & {Full} & Pri-&  Sec- & High- & \\
{Class} & {Sample} & {mary} & {ondary} & {cadence} & {ZTF}\\
\hline
P    & 1.0   & 2.0   & 0.5  & 6.7   & 14.5      \\
QPS  & 33.7  & 24.0  & 39.0 & 28.3  & 35.6      \\
QPD  & 19.3   & 29.4 & 13.9  & 16.1  & 13.8    \\
APD  & 3.9    & 8.4  & 1.5  & 15.7  & 10.0    \\
B    & 18.6   & 15.3 & 20.4  & 17.6  & 8.9   \\
S    & 5.3    & 6.1  & 4.8  & 12.8  &  14.0   \\
EB   & 0.5    & 1.0  & 0.2  &  0.2    & -       \\
L    & 17.6   & 13.8 & 19.7 &2.6   & 3.2\tablefootmark{b}    \\
\hline
\end{tabular}
\tablefoot{\tablefoottext{a}{The reported incidences from previous studies refer to the primary variability classifications and were recalculated to account for the total number of variables in P, QPS, QPD, QPD, B, S, EB, and L.}
\tablefoottext{b}{The L incidence reported for ZTF (3.2\%) does not account for sources classified as such in their secondary classifications ($\sim10\%$).}
  }
\label{tab:comp_gdr3_previous_studies}
\end{table}

Discussion of results must acknowledge survey-to-survey differences. First, in terms of sample size: our study included 9\,824 YSOs in the primary and 18,355 in the secondary sample, while ZTF and high-cadence surveys included 1\,031 and 549 sources, respectively. With much smaller sample sizes, previous studies relied heavily on visual inspection of light curves for classification, with Q-M thresholds adjusted by eye to maximise correspondence with visually assigned morphological classes. Although infeasible in large-scale surveys such as Gaia, this empirical approach adopted in previous studies suggests that survey-to-survey adjustments of Q-M thresholds can help account for surveys' sampling limitations and may improve purity within morphological classes. In Sect.~\ref{subsection_morpho_class}, we adopted \citet{Cody_14} original Q-M thresholds for classification. As further discussed in Sect.~\ref{sec:QMclasses}, our results suggest that these thresholds led to some contamination of symmetric light curves into asymmetric classes within our sample. With the ground truth of visual inspection further hindered by the sparse light curves of Gaia, an alternative approach would be to review Q-M thresholds based on simulated light curves. Although beyond the scope of the present study, our team has been exploring this approach as part of our preparations for Gaia DR4.

High-cadence studies typically have photometric precision of $\sim 0.001$ mag, compared to $\sim 0.2$ mag for ZTF, implying that the ZTF sample is biased towards higher-amplitude variables. In contrast, our GDR3 sample has typical photometric uncertainties $\approx 0.001-0.005$ mag for $G<17$ mag, and $\lesssim 0.02$ mag up to $G\sim19.5$ mag. Based solely on photometric precision, our primary sample should most closely match ZTF, while for $G\lesssim17$ mag, the full sample should best resemble high-cadence surveys. The incidence of morphological classes for the former is shown in Table ~\ref{tab:comp_gdr3_previous_studies}, the latter is omitted, as the results were roughly equivalent to the full sample. However, as discussed in this section, we speculate that cadence and sample selection play a stronger role in explaining different results at the population level across surveys. 

There are also source selection differences, where high-cadence surveys examined only disc-bearing sources, while ours and ZTF results include all YSO types. Moreover, previous studies focused on individual star-forming regions, with much narrower age distributions in comparison with the GDR3 all-sky sample. Notably, \citet{Cody_2022} attempted to interpret the frequency distribution of morphological classes in different star-forming regions in light of their age. In our case, this type of interpretation is hindered by the lack of previous constraints in the age distribution of the GDR3 YSO catalogue. However, our investigations of $\alpha_\mathrm{IR}$-indices in Sect.~\ref{sec:alpha_ir} revealed that 42\% and $\sim 32\%$ of our sample of variables had no indication of the presence of a disc or only a thin-disc contribution to their SEDs. 

Given the enhanced fraction of discless YSOs in our sample, cool spot variability, typically associated with P and QPS variability, is expected to be the most common in our sample. Although QPS is in fact the most common variability class,  Table~\ref{tab:comp_gdr3_previous_studies} shows that P frequency is much lower than in previous studies. Beyond cadence and underlying noise, we further discuss in Sect.~\ref{sec:P} that this can be attributed to cool-spot variability being unstable at timescales of the order of GDR3 observations. We note that cool-spot variability is a central feature of ``solar-like variables'', which are themselves a subgroup of GDR3 variability classes, with \citet{Distefano2022} reporting that 53--72\% of their 474\,026 variables show unstable spot-variability (amplitude variation) during the observed time covered by GDR3. This helps explain not only our under-detection of P variables compared to previous studies, but also contributes to our enhanced detection of QPS variables.  Table~\ref{tab:comp_gdr3_previous_studies} also indicates that ZTF studies, which also include discless sources, show a much larger sample of P and QPS sources in comparison to high-cadence surveys. However, ZTF studies use much larger Q thresholds to account for the higher noise from ground-based observations. 

The solar-like variables reported by \citet{Distefano2022} focus on late-type stars, including some relatively young objects, while being disjoint from the GDR3 YSO sample. This suggests that a certain fraction of young P and QPS variables may be lost to their sample. To quantify sources lost to the GDR3 sample of solar-like variables, we further examined the NEMESIS YSO catalogue for the Orion Star Formation Complex \citep{roquette2024}. Of 27\,879 YSO candidates, the NEMESIS sample included 5\,231 sources in common with the Gaia YSO sample, of which 111 were classified as P and 2\,131 as QPS, and a further 152 sources in common with Gaia solar-like variables, suggesting that we miss about $\sim7\%$ of cool-spot variables. Similarly, EBs are also a subclass of variables in Gaia, with \citet{Mowlavi2023A&A...674A..16M} reporting more than 2 million EB candidates as part of DR3. A similar comparison to the NEMESIS catalogue suggests that our EB detection rate might be reduced by half. 

We also point to survey-to-survey differences in the time span of observations. High-cadence surveys cover between 40 and 80 days of observations, while ZTF surveys cover 800--1\,500 days, and GDR3 750--1\,000 days. These differences imply that quasi-periodic phenomena at intermediate timescales may be perceived as aperiodic by studies with reduced time span. Moreover, L variables are defined as sources with timescales larger than 20-40 d in high-cadence studies, 100-250 d in ZTF, and we used 350-500 d in the present study. L variable fractions are thus fully survey-dependent. Even if removed from our class incidence comparison, their impact on the definition of other morphological classes is still worth mentioning. First, L variables could be operationally defined as aperiodic, since their time scale was not resolvable over the duration of a survey. Although their lack of reliable timescale implies that their Q indices cannot be estimated, accounting for their M indices would increase class frequencies by 3.8\% for B, 8.6\% for S, and 5.3\% for APDs - thus bringing our class incidence for S and APD variables to levels similar to those of ZTF studies. Second, the timescale limit for the L variable definition indirectly determines how many folds are used in the waveform estimation required to derive the Q index. With our threshold of half of the total duration of observations, only one fold is included for sources with the longest timescales, which may lead to over-fitting of estimated waveforms, resulting in deflated Q indices, and causing aperiodic phenomena to be miss-detected as quasi-periodic. 

Some previous studies attributed multiple morphological classes to multi-mode variables \citep[e.g.][]{Cody_2018,Cody_2022,Hillenbrand2022} also including an additional class, multi-periodic, to account for variables with multiple periods. In Table~\ref{tab:comp_gdr3_previous_studies}, multi-mode variables were represented by their primary morphological class. However, with our focus on the dominant variability mode, we acknowledge that undetected multi-mode variables in our sample may display inflated Q indices. Beyond multiple variability mechanisms occurring simultaneously, the longer time span of surveys like Gaia and ZTF may also facilitate the observation of sources with alternating dominant variability phenomena. For example, this can result in symmetric M indices derived from light curves that shift from bursting to dipping behaviour (e.g., Fig.~\ref{fig_lc_s}).

Our total fraction of dipping variables (QPD and APD) matches that reported in ZTF studies, but is lower than in high-cadence surveys. Nevertheless, as already noted, our APD fractions are underestimated due to our definition of L variables. Furthermore, in Sects.~\ref{sec:QPD} and \ref{sec:APD} we discuss evidence that our sample of dippers includes a different underlying population of YSOs in comparison to the high-cadence surveys. Similarly, B variables also include different variability mechanisms, and in Sect.~\ref{sec:disc_bursters} we discuss evidence that the Gaia sampling yields the detection of a distinct population of bursting variables in comparison with other surveys.

\subsection{General properties of the morphological classes}\label{sec:QMclasses}

In this section, each variability morphological class is discussed within the context of its expected physical origins and its typical timescale and amplitude. 

\subsubsection{Strictly periodic}\label{sec:P}

\begin{figure}[h!]
   \centering
      \includegraphics[width=1\linewidth]{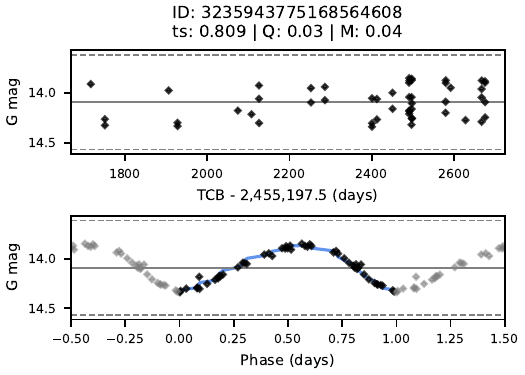}
      \caption{Example of a P variable. Top: Full $G$ light curve. Bottom: Phase-folded light curve. The blue line represents the waveform used in the Q calculation, the grey line indicates the mean magnitude, and the dashed lines mark the $3\sigma$ variation bounds. Same descriptions applies to all following figures until Fig.~\ref{fig_lc_l}.}
         \label{fig_lc_p}
   \end{figure}

Strictly periodic variables (P, Fig.~\ref{fig_lc_p}) are usually associated with rotational modulation by magnetically induced cool spots at the stellar surface in sources viewed close to edge-on \citep[$\sim40-90^{\circ}$;][]{Cody_2022}. Fig.~\ref{fig:Q-M_boxplots_comp} shows that the timescale distribution for P sources is consistent across studies and dominated by rotational timescales. The variability amplitude in our primary sample (blue) is higher than that reported in high-cadence surveys (red) but slightly lower than in ZTF studies (orange); the secondary sample (grey) exhibits amplitude levels comparable to those of high-cadence studies. Differences in amplitude distributions can be attributed to selection effects in different surveys. 
Typically, cool spots have lifetimes ranging from 1 to 50 rotation cycles \citep{Basri2022ApJ...924...31B}. Higher-cadence surveys, which usually span 40 to 80 d, provide excellent phase coverage before significant changes in spot coverage occur. In Gaia, sparse sampling and long-term coverage implies that hundreds of rotation cycles take place before achieving sufficient phase coverage, thus biasing the sample of P sources towards sources with larger and more stable spot coverage (and large amplitudes). In contrast, the higher noise level of ground-based observations allows only the higher-amplitude variables to be detected. As a result, Fig.~\ref{fig:Q-M_boxplots_comp} shows that the ZTF distributions of all classes are shifted toward higher amplitudes compared to other surveys. 

The properties of our P variables support the hypothesis that their variability is primarily driven by rotational modulation linked to relatively stable cool spots. Their average amplitude is among the lowest of all morphological classes, and the ratio between rotation and intermediate timescales (Table~\ref{tab:gdr3_allsky}) shows that this sample is dominated by rotational modulation. P variables are 82.95\% composed of YSOs without evidence of a thick disc or envelope, and 90.2\% composed of K- and M-type stars. Furthermore, 86.27\% of P sources with H$\alpha$ information are WTTS.

\subsubsection{Quasi-periodic symmetric}\label{sec:QPS}

\begin{figure}[htb]
   \centering
      \includegraphics[width=1\linewidth]{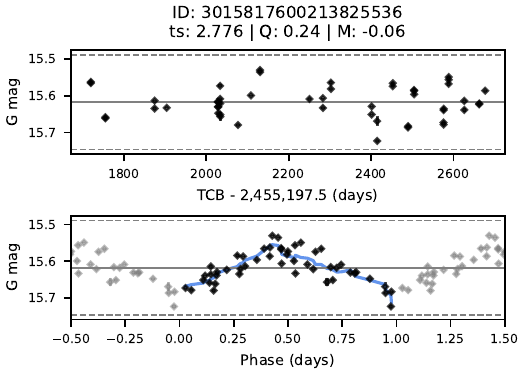}
      \caption{Example of a QPS variable.}
         \label{fig_lc_qps}
   \end{figure}

 Multiple physical origins have been proposed to explain QPS variables (Fig.~\ref{fig_lc_qps}), which are the most common in our entire sample. Although generally linked to rotational modulation of stellar spots, high-amplitude QPS have been tentatively associated with variability caused by large hot spots induced by accretion, in combination with underlying lower-amplitude accretion variation \citep{Venuti2015A&A...581A..66V,Cody_2018}, while low-amplitude QPS are thought to originate from changes of cool spot coverage due to magnetic activity \citep{Cody_2022}. Considering that cool spots are the only variability driver expected in discless sources and that disc-bearing and actively accreting sources are likely to exhibit accretion-induced hot spots, this view is generally supported by our sample. We find that the average distribution of amplitudes is lower for discless and thin-disc QPS (0.11 mag) compared to sources with envelope or thick disc (0.30). Similar differences are found when comparing the amplitudes of the WTTS QPS (0.11) with the CTTS QPS (0.36). The same trend is verified by comparing our primary and secondary samples, where the CTTS frequency varies from 36.6\% to 6.1\%. The link of QPS with rotational modulation by spots is further reinforced by the observation that M-type QPS have an excess of short timescales, with 60\% of them displaying timescales of $\lesssim 5 $ days compared to $\sim 30\%$ for other spectral types. This is in accordance with the expectation that M-type stars tend to be faster rotators than their more massive counterparts \citep[e.g.][]{2007ApJ...671..605C}.

Our QPS-variables include a population of sources with intermediate timescales, which has not been found in previous studies. Fig.~\ref{fig:Q-M_boxplots_comp} shows that for the primary sample, the timescale distribution is bimodal, with 1.3 times as many rotation timescales as intermediate ones. A comparison of their Q index distributions reveals significant differences (with KS test support), where the rotational-timescale QPS are shifted toward higher periodicity with an average Q of 0.34 compared to 0.44 for intermediate-timescale QPS. 

Visual inspection of intermediate-timescale QPS light curves revealed that many of these sources have limited phase coverage when folded to their characteristic timescale. This limitation is inherent in surveys with sparse sampling as Gaia, where some phase-folded light curves include large gaps with no data. These gaps, in turn, bring us to the limits of the method for waveform estimation, causing waveforms to be over-fit to only a fraction of the phase, artificially shifting Q indices towards lower values and causing misdetection of aperiodic sources as quasi-periodic. With the Gaia scanning law, the timescales that yield gaps in phase coverage vary with sky position, making modelling and masking these timescales from our analysis impractical. To mitigate this issue, future studies should consider the complementation of Q-M analysis with an additional variability metric sensitive to the degree of phase coverage as part of the procedure for the derivation of the Q index \citep[e.g.][]{Saunders2006AN....327..783S}. We also considered adjusting the quasi-periodic/aperiodic Q index threshold to mitigate this effect, but verified that a lower Q index limit would result in reliable quasi-periodic light curves with rotation timescales being misclassified as aperiodic. 

\subsubsection{Quasi-periodic dippers}\label{sec:QPD}

\begin{figure}[ht]
   \centering
      \includegraphics[width=1\linewidth]{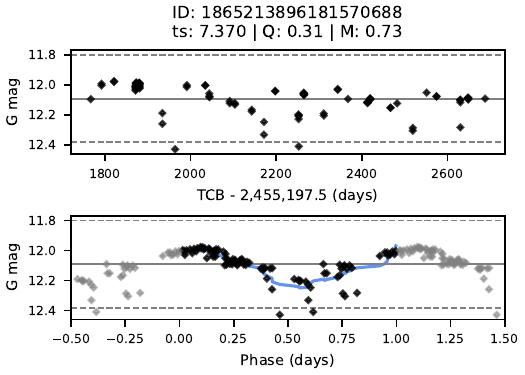}      \includegraphics[width=1\linewidth]{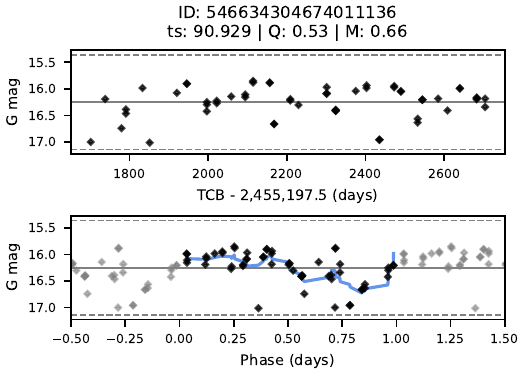}      \includegraphics[width=1\linewidth]{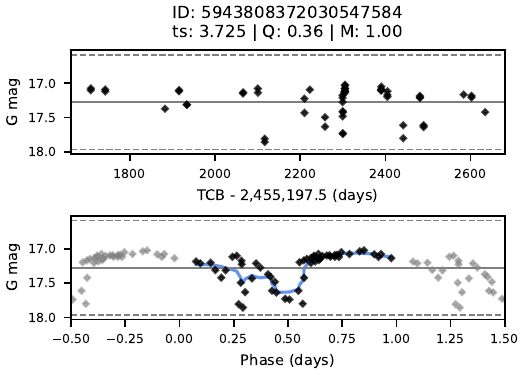}
      \caption{Examples of QPD variables at rotational timescales (top and bottom) and intermediate timescale (middle).}
         \label{fig_lc_qpd}
   \end{figure}

The phenomenology AA Tauri variables \citep{Bouvier1999A&A...349..619B} is often linked to QPD light curves (see Fig.~\ref{fig_lc_qpd}), where a misalignment between the rotation and magnetic field axes yields the formation of wraps, with disc material lifted above the disc mid-plane and close to the co-rotation radius, introducing quasi-periodic occultations at rotational timescales in systems observed close to edge-on \citep{Romanova2013MNRAS.430..699R,McGinnis2015A&A...577A..11M}. 
Among the primary sample, 62.09\% of QPDs have an inner disc, and 40.60\% have evidence of ongoing accretion (CTTS classification), supporting the association of this class with physical phenomenology related to the presence of a disc and ongoing accretion. However, a direct association of these with the AA Tau prototype is hampered by two factors. 

Firstly, a fraction of our low-amplitude QPDs may be misclassified QPS. About half of our QPDs are in the secondary sample, which shows a much lower disc (13.71\%) and CTTS (4.47\%) frequencies, with both Q and M distributions shifted towards more symmetric M and more periodic Q indices compared to the primary sample. Fig.~\ref{fig:Q-M_boxplots_comp} shows that these have systematically lower amplitudes than QPDs from high-cadence surveys. The M index distribution for the secondary QPD sample peaks near the threshold for dipper selection, with 90\% of the M indices below the median value for the primary QPD sample. Additionally, although statistically different in shape, the distributions of amplitudes and Q index for the secondary QPD and secondary QPS samples have similar averages and standard deviations, as well as similar disc and CTTS frequency. Finally, visual examination of the light curves also suggested that a higher M index threshold might be necessary to enhance the purity of the dipper variable samples, where a reviewed threshold would identify a large fraction of secondary QPD as QPS.

Secondly, because the observation time span of Gaia is much longer, QPDs may be sensitive to other dimming variables beyond AA Tau. Fig.~\ref{fig:Q-M_boxplots_comp} reveals a bimodal distribution for QPDs, with $\sim35\%$ of QPDs showing intermediate timescales. This contradicts the predictions of the AA Tau phenomenology, where the occulting material is anticipated to be near the co-rotation radius. The primary sample of QPDs with intermediate timescales shows high disc and CTTS frequencies (65.09, and 47.35\%), suggesting a physical origin also associated with the presence of a disc and ongoing accretion. Statistical tests for distribution shape, mean and variance differences for the distributions of Q indices and amplitude support the hypothesis that our primary QPD sample is composed of two distinct populations of variables with rotation and intermediate timescale, where the rotation timescale QPDs have lower amplitude on average (0.29 mag) and are more periodic (lower Q), while the intermediate timescale QPDs tend to have larger amplitude on average (0.44 mag) and reduced periodicity in comparison. We note that this second population of intermediate-timescale QPDs is also present in the ZTF sample, which has a time span close to that of Gaia, suggesting that this second type of QPDs is absent from high-cadence surveys due to their limited total time span.

While QPDs at rotational timescales can be associated with AA Tau type variability, those with intermediate timescales could be related to occultation by disc material located farther away than the inner disc. One possible prototype for this variability behaviour could be the young star RW Aur \citep{Cabrit2006A&A...452..897C, Facchini2016A&A...596A..38F}. A proposed phenomenological explanation for the fading of RW Aur would be a tidally truncated disc in which a flyby removed the outer part of the disc, leaving a tidally disrupted 'arm' that induces occultations of stellar light when observed from moderate viewing angles \citep[e.g.][]{Rodriguez2013AJ....146..112R}. We note that although RW Aur is best known for its 2 mag deep fading events in 2010 and 2014, lower amplitude variability could be expected under the same phenomenology \citep[e.g][]{Fischer23}. A different possible phenomenon could be disc instabilities as a trigger of dense disc winds responsible for enhancing line-of-sight extinction above the disc mid-plane, which, due to their radial location, would cause orbital modulation at longer timescales \citep[e.g.][]{Bouvier2013A&A...557A..77B}. Further constraints on the colour behaviour during dipping features could help further probe QPDs.

\subsubsection{Aperiodic dippers}\label{sec:APD}

\begin{figure}[htb]
   \centering
      \includegraphics[width=1\linewidth]{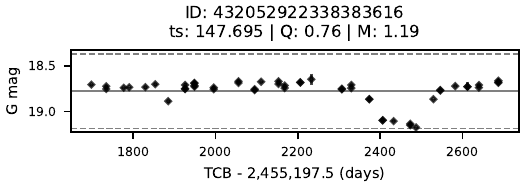}
      \includegraphics[width=1\linewidth]{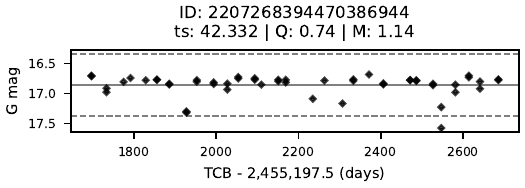}
      \includegraphics[width=1\linewidth]{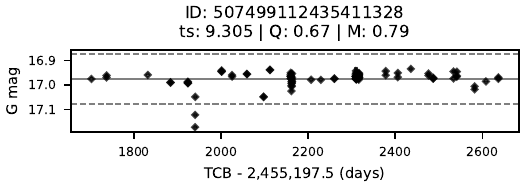}
      \caption{Examples of APD variables.}
         \label{fig_lc_apd}
   \end{figure}

With the highest inner disc and CTTS frequencies (56.94 and 39.64\%), APD variability (Fig.~\ref{fig_lc_apd}) is likely associated with the star-disc interaction process. Previous studies have also linked APDs to the physical phenomenology of AA Tau. However, while QPDs require star-disc systems that undergo stable accretion to allow dust to be lifted above the disc midplane long enough for the occultations to be observed as quasi-periodic, in APDs, an unstable accretion regime will instead result in stochastically distributed aperiodic dippers through the light curve \citep[][]{KulkarniRomanova2008MNRAS.386..673K,McGinnis2015A&A...577A..11M}. Nevertheless, material occulting the star is still expected to be at around the co-rotation radius, thus aperiodic occultation features are still expected to have durations around rotational timescales. In contrast, Tables~\ref{tab:gdr3_allsky} and Fig.~\ref{fig:Q-M_boxplots_comp} reveal that APDs show intermediate timescales 2.26 times more often than rotational ones. 

We note that a meaningful interpretation of aperiodic timescales can be tricky. 
Fig.~\ref{fig:Q-M_boxplots_comp} shows that APDs from high-cadence surveys have timescales between a week and a month, with these surveys being limited to a time span of 40-80 days. { If their APDs are} in reality an undersampled quasi-periodic phenomenon occurring over timescales longer than their time span, these will be detected as quasi-periodic by surveys like Gaia or ZTF, which explains the population of intermediate timescales among our QPDs, and the reduced population of rotation timescales in our primary APDs sample. Second, most aperiodic light curves had their timescale determined by the SF in Sect.~\ref{sf_ts}. SF aperiodic timescales measure the time it takes for the dominant variability feature in the light curve to reach its maximum amplitude, thus reflecting the duration of the variability feature, rather than the frequency of occurrence of variability over the span of observations. For example, the bottom panel of Fig.~\ref{fig_lc_apd} shows an APD light curve potentially linked to the AA Tau phenomena, where the 9.3 d derived aperiodic timescale reflects the time to reach the maximum amplitude (0.24 mag) in the main dip visible in the light curve. In the top panel of the same figure, we show a second example where the timescale of 147.7 days reflects half of the duration of a main fading event with amplitude 0.53 mag. Combined with the large amplitudes found among our primary sample of APDs, the large disc and CTTS frequencies and large aperiodic timescales support the suggestions that beyond AA Tau-like light curves, APDs at intermediate timescales may be a population of moderate-amplitude RW Aur fading light curves \citep[e.g.][]{Petrov2015}. 

With 8.3\% of APDs having estimated spectral types between AB and F and 87\% of these having a disc, a possible explanation for these earlier type APDs could be the phenomenology behind UX Ori variables \citep{Natta1997ApJ...491..885N}. These YSOs have their variability attributed to large-scale disc perturbations causing a puffing-up of disc material close to the dust sublimation radius \citep{2022MNRAS.512.3098S}, which, when observed almost edge-on, results in sporadic and large amplitude (1--2 mag) fades. Nevertheless, we verified that although the distributions of amplitudes of earlier and later type APDs are statistically different, the former have much lower typical amplitudes ($0.36\pm0.04$ mag) compared to the latter ($0.54\pm0.02$ mag).

\subsubsection{Bursters} \label{sec:disc_bursters}

Based on correlations with various accretion proxies, from UV excess to variability in colour behaviour and H$\alpha$, B variables were previously associated with high levels of accretion \citep{Stauffer2014AJ....147...83S,Cody2017,Venuti_2021,Hillenbrand2022}. In our sample, this association with star-disc-interaction phenomenology is only supported for sources in our primary sample, for which the inner disc and CTTS frequency were 45.18\% and 45.49\%, while for the secondary sample these are 12.22 and 7.04\% respectively. Among these with evidence of a disc or ongoing accretion, we note that the earlier and later type sources share statistically different distributions of amplitude, with the distribution of earlier types (mean: $0.57\pm0.06$ mag) shifted towards higher amplitudes in comparison with later type sources ($0.38\pm0.02$ mag).

\begin{figure}[htb]
   \centering
      \includegraphics[width=1\linewidth]{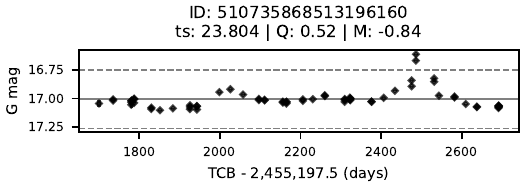}
      \includegraphics[width=1\linewidth]{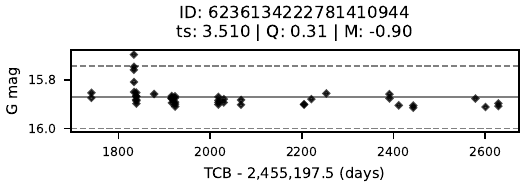}
       \includegraphics[width=1\linewidth]{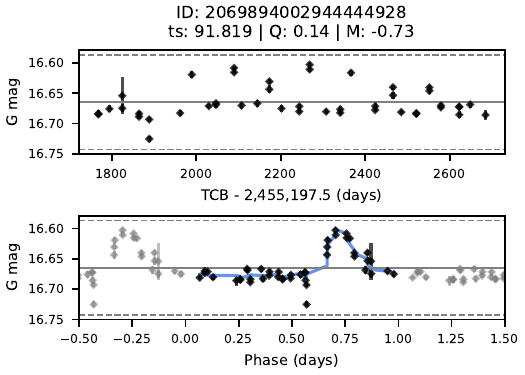}
      \caption{Examples of B variables with an intermediate duration burst (top), a short duration burst coinciding with a cusp (middle) and a quasi-periodic bursting behaviour (bottom, also shown phase-folded).}
         \label{fig_lc_b}
   \end{figure}

As defined by \citet{Cody_14}, the B class might constitute a family of bursting behaviours, rather than relating to a single physical mechanism. In a study focused solely on B variables, \citet{Cody2017} finds varied bursting behaviour, with some sources showing a continuous series of bursts, while others presenting short-lived burst episodes on top of underlying irregular variability \citep[see also][]{Venuti_2021}. This family of bursting behaviours is verified among our primary sample, and illustrated in Fig.~\ref{fig_lc_b}, where we show examples of isolated bursts, including both short-lived (GDR3 510735868513196160) and longer-lived bursts (GDR3 6236134222781410944). Fig.~\ref{fig_lc_b} also includes an example of a quasi-periodic B variable with intermediate timescale, GDR3 2069894002944444928, which may be related to the phenomenology behind the DQ Tau system \citep{Tofflemire2017ApJ...835....8T}, where two stars orbit each other in a highly eccentric orbit and their magnetic interaction with circumbinary and circumstellar material yields enhanced accretion near periastron, resulting in the observation of regular bursts. Following the discussion on outlier removal in Sect.~\ref{sec:spurious_outliers} and App.~\ref{app:outlier}, we note that the sparse cadence of Gaia is expected to yield an under-detection of variables with short-lived bursts. As reported by \citet{Cody2017}, only 1 burst out of 29 monitored with high-cadence K2 light curves showed burst durations longer than 10 days. If observed by Gaia, such a sample would likely have burst events ruled out as outliers, unless the bursting feature coincides with a cusp, as in the case of a short-lived burst variable GDR3 6236134222781410944 shown in Fig.~\ref{fig_lc_b}. Finally, we did not find examples of extreme burst events, such as a FUor or EXors. However, we verified the existence of some sharp long-term bursts among our long-trend variables (e. g.,  Fig.~\ref{fig_lc_l}).

\subsubsection{Stochastics}\label{sec:S}

\begin{figure}[htb]
   \centering
      \includegraphics[width=1\linewidth]{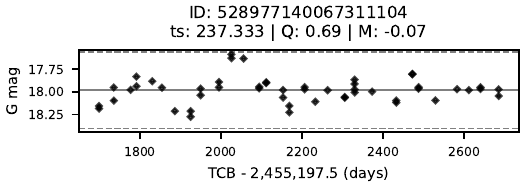}
      \includegraphics[width=1\linewidth]{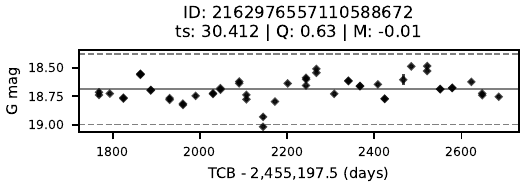}      
      \caption{Examples of S variables without preference for bursting or dipping (top) and  alternating between dipping and bursting (bottom).}
         \label{fig_lc_s}
   \end{figure}

The light curves of S variables (Fig.~\ref{fig_lc_s}) are dominated by irregular variability without detectable periodicity or a clear preference for bursting or dimming behaviour.
Fig.~\ref{fig:Q-M_boxplots_comp} shows that even our primary sample shows systematically lower amplitudes compared to previous studies. 
Previous high-cadence surveys associated S-variables with a combination of the erratic routine variability introduced by accretion, combined with viewing angles close to face-on, which hindered the rotational modulation of accretion-induced variability \citep{Stauffer2016AJ....151...60S}. This scenario is supported by our primary S sample, which has inner disc and CTTS frequencies of 43.26 and 49.04\%, respectively, along with 2.3 times as many intermediate timescales as rotational ones. However, visual inspection of S light curves shows that a fraction of those are multi-mode variables with alternating bursting and dipping behaviour across the observed time (see Fig.~\ref{fig_lc_s} bottom panel). In contrast, our secondary sample has very low inner disc and CTTS frequency (18 and 7.36\%), with a wide spread of timescales and no clear preference for rotational or intermediate timescales. However, when disaggregated by amplitude, this sample reveals an excess of sources with low-amplitude (below 0.1 mag) fast rotation timescales (below 5 days), which may suggest a link between this sample and cool spot-dominated light curves close to the Gaia detection limit. 

\subsubsection{Eclipsing binaries}\label{sec:EB}

\begin{figure}[htb]
   \centering
      \includegraphics[width=1\linewidth]{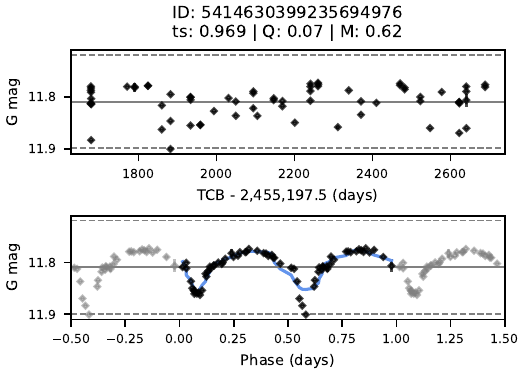}
      \includegraphics[width=1\linewidth]{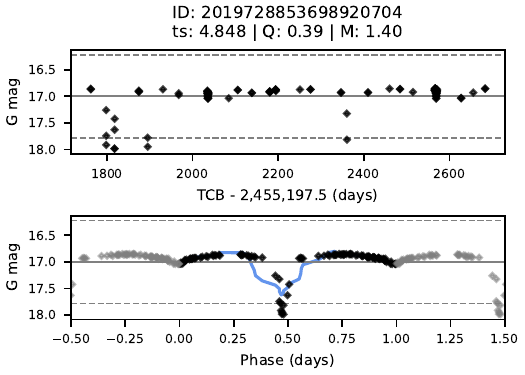}
      \caption{Examples of EB variables, including a QPD misclassification (bottom) due to timescale detection of a harmonic instead of the true period. The phased light curves are shown folded to their true period. }
         \label{fig_lc_eb}
   \end{figure}

\begin{figure}[htb]
   \centering
      \includegraphics[width=1\linewidth]{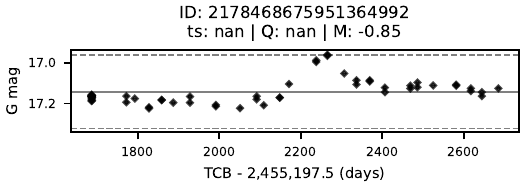}
      \includegraphics[width=1\linewidth]{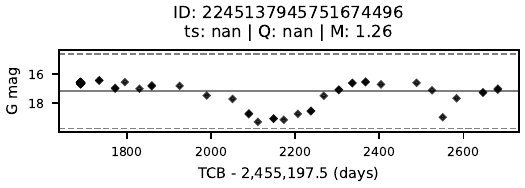}
      \includegraphics[width=1\linewidth]{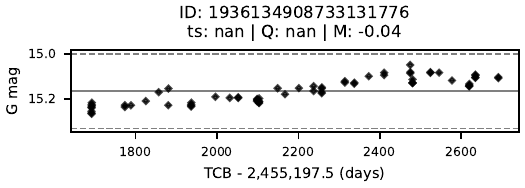}
      \caption{Examples of L variables for a bursting (top), dipping (middle), and symmetric (bottom) source.}
         \label{fig_lc_l}
   \end{figure}

The EB class in \citet{Cody_14} targeted prototypical eclipsing binary light curves with strictly periodic asymmetric light curves. We found 129 EB sources. A search for counterparts in the Simbad Database \citep{2000A&AS..143....9W} reveals that only 4 of our EB sources were previously identified as such in the literature. Although sources classified as EB generally followed the prototype of strictly periodic eclipses, our automated method for timescale derivation using the GLS (Sect.~\ref{gls_ts}) often detects harmonics of the true period as the best timescale. We show two examples of EB light curves in Fig.~\ref{fig_lc_eb}, where in both cases the GLS returned half of the true period as the best timescale. In the case of the source GDR3 5414630399235694976 (top plots), secondary and primary eclipses are relatively similar in depth, with the waveform estimated based on the biased timescale representing the true waveform relatively well and yielding a very low Q value. However, in cases like the source GDR3 2019728853698920704 (bottom plots), as the secondary eclipse is much shallower, waveform estimation based on the harmonic of the true timescale inflates Q values, resulting in the miss-classification of the source as a QPD.

\subsubsection{Long trends}\label{sec:L}

With timescales long for a relevant waveform to be derived from the phase-folded curve, L variables lack a Q index, but we still computed their M index. In the primary sample, 21\% Ls have bursting behaviour and 30\% have dipping behaviour. For the secondary sample, 21\% Ls have bursting behaviour, and 26\% have dipping behaviour. L variables tend to have larger amplitudes than the other classes, which is expected since there is an empirical correlation between the timescale of an event and its amplitude \citep[See Fig. 3 of][]{Fischer23}. Although we do not verify any preference for bursting or dipping as a function of their spectral types, we find that earlier type sources have typically much lower amplitudes (mean: $0.095\pm0.004$ mag) compared with later type sources ($0.187\pm0.005$ mag).

 \subsection{Comparison across surveys}\label{subsec:comparison_previous_studies}

To further investigate the consistency of Q and M indices across different surveys, we cross-matched our variable sample with previous Q-M studies using K2 \citep{Cody_2022} and ZTF \citep{Hillenbrand2022, Wang2023, JiangHillenbrand2024}. Figs.~\ref{fig:lc_gdr3_k2} and \ref{fig:lc_gdr3_ztf} show examples of light curves of sources with observations in multiple surveys. Q-M indices from the literature are adopted as reported in the source papers. None of these surveys included seasons of observations in common.

\subsubsection{High-cadence space-based surveys}

\begin{figure}[htb]
        \includegraphics[width=.9\linewidth]{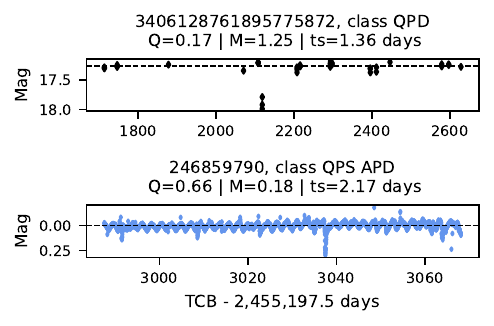}
        \includegraphics[width=.9\linewidth]{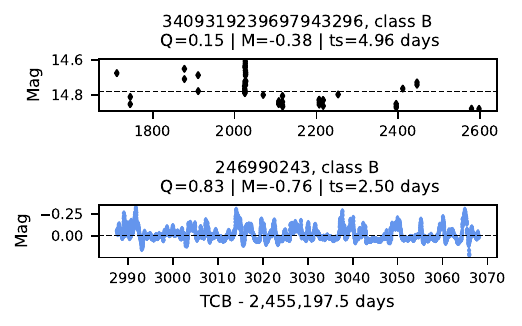}

    \caption{Comparison between GDR3 (top) and K2 (bottom) light curves. The horizontal dashed black lines show the median magnitude, whereas average magnitudes were subtracted from the K2 light curves.}
    \label{fig:lc_gdr3_k2}
\end{figure}

We found only 13 sources in common between our variable sample and the sources observed by \citet{Cody_2022} with K2 in Taurus (see Fig. \ref{fig:lc_gdr3_k2}). Hence, we limit our comparison to a qualitative discussion.
Since K2 and GDR3 have very different cadences and time spans, our comparison focuses on their M index behaviour, where classification was considered equivalent if the source was detected as dipping, symmetric or bursting in both surveys. Although variability amplitudes were generally consistent across surveys, Q-M indices often had large enough differences to produce distinct morphological classes. 

Of the three symmetric sources in K2, only one was also symmetric in GDR3, while the others were dipping. However, since these surveys cover different seasons of observation, it is impossible to infer whether discrepant classifications result from a change in dominant variability mode or the result of different cadences of observations. Three B-variables in K2 were also B in GDR3. One of these sources is GDR3 3409319239697943296 (Fig.~\ref{fig:lc_gdr3_k2}), which, although bursting in both surveys, was quasi-periodic in our study and aperiodic when observed by K2. Of the four dipping variables in K2, three were also dipping in GDR3.  
\citet{Cody_2022} searched for multiple variability modes by computing Q-M twice: first on the original light curve and then on the residual data. Two out of three multi-mode sources in common with GDR3 have one of their morphological classes matched to GDR3. An example is shown in Fig.~\ref{fig:lc_gdr3_k2}, where the QPD source GDR3 3406128761895775872 was attributed to the QPS and APD classes in the K2 analysis. 

\subsubsection{Comparison with ZTF}\label{ground-based}

\begin{figure}[htb]    
        \centering
        \includegraphics[width=\linewidth]{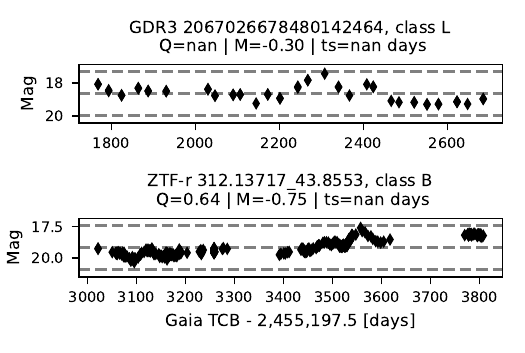}
        \includegraphics[width=\linewidth]{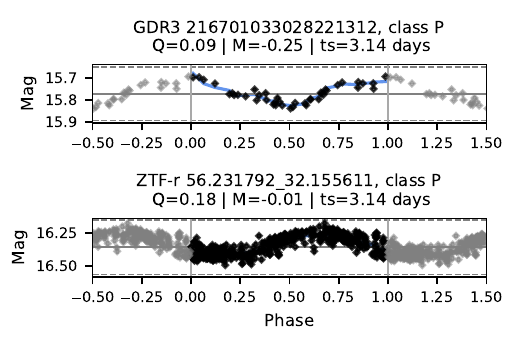}
    \caption{Comparison between GDR3 (top) and ZTF (bottom) light curves. The periodic source (bottom) is phase-folded to its detected timescale in each survey.}
    \label{fig:lc_gdr3_ztf}
\end{figure}

\begin{figure}
   \centering
      \includegraphics[width=.9\linewidth]{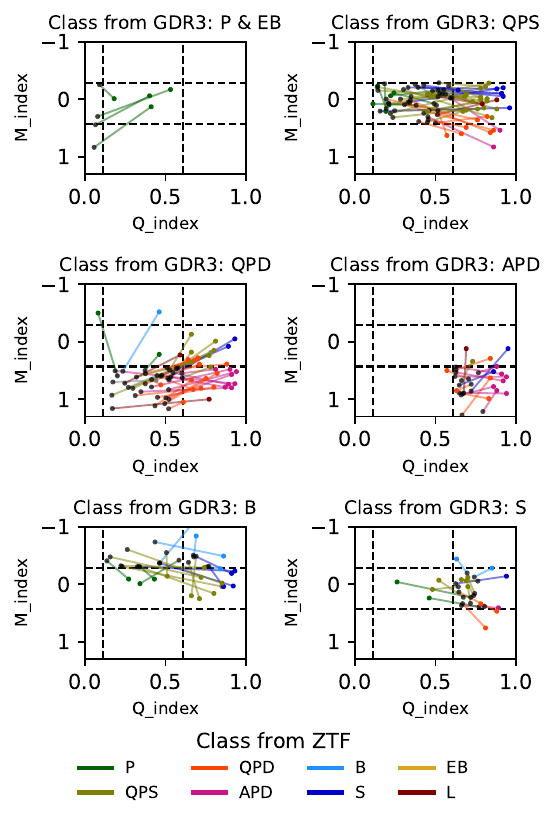}
      \caption{Q-M variation across surveys for sources observed in both GDR3 (black dots) and ZTF (colour dots). Line segments highlight Q-M variation across surveys, with colour matching thier ZTF class. }
         \label{fig_gdr3_ztf_qm}
\end{figure}

Light curves observed by ZTF have roughly the same duration as GDR3 but are noisier due to ground-based observations. With 165 sources in common, Fig.~\ref{fig_gdr3_ztf_qm} compares the Q-M distribution between surveys, with 25 sources omitted as they were classified as L variables (no Q index) in the present survey. At the population level, the two surveys yield broadly consistent results, particularly for light curve symmetry with very few cases where variables were detected in radically different behaviours, for example, switching from bursts to dips. For periodic and quasi-periodic sources, we find that differences in derived timescales were either due to harmonics or to one-day aliasing caused by ZTF ground-based observations. For aperiodic sources, timescales differed substantially, but this was explained by the distinct methods used to derive them. Given the different Q index thresholds adopted in ZTF surveys, most QPS variables in Gaia were also QPS in ZTF. Some dippers in ZTF were identified as QPS or S variables in GDR3, although they are generally close to the M threshold in our study. Most dippers (QPDs or APDs) in Gaia were also dippers in ZTF. Most bursts in Gaia are detected as symmetric in ZTF, although we note that both surveys' low cadence makes them inefficient in detecting bursts. Fig.~\ref{fig:lc_gdr3_ztf} shows two examples of sources in common in both surveys. GDR3 2067026678480142464 (top) shows similar behaviour in both surveys, and although we classified it as an L variable, its M index suggests the same bursting behaviour, with a slight shift towards higher symmetry in GDR3 than in ZTF. GDR3 216701033028221312 (bottom) has a similar behaviour in both surveys, with matching timescales but higher phase dispersion in ZTF  than in GDR3, resulting in a higher Q index value for ZTF. Additionally, the ZTF light curve contains significantly more data points, contributing to its more symmetric behaviour compared to GDR3.

\section{Summary and conclusions}\label{section_conclusion}

We presented the results of a pilot study employing the asymmetry and periodicity variability metrics to characterise an all-sky sample of YSO candidates based on their G-band photometric variability observed with Gaia. We adapted the method behind the Q-M indices for application to Gaia's long-term but sparse light curves. Starting from the sample of 79\,375 variables identified in GDR3 as YSO candidates based on their variability, we pre-processed sources to filter for a low number of epochs, remove contaminants, and identify sources that were likely affected by spurious instrumental variability. Based on a sample of 28\,179 sources, we inferred characteristic variability timescales by considering the possibility of periodic and aperiodic variability. The timescale distribution revealed a bimodality with rotational and intermediate timescales, with the former showing,  on average, lower variability amplitudes than the latter. Rotational timescales were more frequent for sources with a higher periodicity and intermediate timescales for aperiodic sources. We successfully derived Q-M indices for 23\,213 sources. 
We complemented our analysis with observations of H$\alpha$ as a proxy of ongoing accretion and $\alpha_{IR}$ as a proxy of the presence of circumstellar material and the evolution stage of YSOs. 
We found that WTTS discless sources tend to show more periodic and symmetric variability, while CTTS and WTTS disc-bearing sources tend to be more asymmetric and less periodic. We also examined the results as a function of the estimated spectral types for the sample and found that QPS variability is much more common among M type stars, while the incidence of L variability increases towards earlier types.

We used the derived Q-M indices to sort sources into one of six variability morphology classes: Strictly periodic, quasi-periodic symmetric, stochastic, quasi-periodic dipper, aperiodic dipper, burst, or eclipsing binary. The timescales of 4\,966 additional sources were longer than half of the total observed time, and these were classified as long trend variables with only their M index derived. 
The subsamples in each of these classes were further examined in terms of Q-M indices, variability amplitudes, timescales, and accretion and disc proxies, and we discussed possible links to well-known variability mechanisms. This examination revealed a wider variety of variability behaviour than previous studies, which we attribute to the longer time span of Gaia. We found sources displaying the usual variability caused by rotational modulation by magnetically induced spots or accretion-induced hot spots, and both bursting and dipping behaviour at rotational timescales, likely linked to star-disc interaction. Moreover, we found a relevant fraction of sources showing dipping and bursting behaviour at an intermediate timescale, which we tentatively linked to the phenomenology behind RW Aur fades and DQ Tau-type bursting variability. 

We qualitatively compared our Q-M-based results with previous studies, including both high-cadence (CoRoT and K2) and ground-based surveys (ZTF), and discussed them at the individual and population levels. Although large survey-to-survey differences were found, the M index trends seem generally consistent. When considering survey-specific biases (caused by cadence, time span, precision, and age), we found that differences in the incidence of morphological classes across surveys can be explained.
We note a lack of available observed datasets with common sources observed by Gaia and high-cadence surveys, which would allow us to probe the appropriateness of Q-M thresholds for a classification across surveys. 
Future studies should consider assessing the effect of survey-to-survey variations on these thresholds based on simulated data. Finally, Gaia DR4 will allow us to expand these results to an even broader all-sky coverage because the time span of the observations is longer and the calibration is improved.

\begin{acknowledgements}

We acknowledge funding from the European Union’s Horizon 2020 research and innovation program (grant agreement No.101004141, NEMESIS).
G.M. acknowledges support from the János Bolyai Research Scholarship of the Hungarian Academy of Sciences. 
We thank Berry Holl for helpful discussions on the use of Gaia DR3 data, Ann-Marie Cody and Laura Venuti for exchanges about the implementation of the Asymmetry and Periodicity metrics. 
This research was also supported by the International Space Science Institute (ISSI) in Bern, through ISSI International Team project 521 selected in 2021, Revisiting Star Formation in the Era of Big Data (\url{https://teams.issibern.ch/starformation/})
This work made use of Astropy \citep{astropy:2013, astropy:2018, astropy:2022}, Scikit-learn: Machine Learning in Python \citep{scikit-learn}, Pandas \citep{reback2020pandas}, Numpy \citep{harris2020array},  and Matplotlib \citep{Hunter:2007}.
This work has made use of data from the European Space Agency (ESA) mission
{\it Gaia} (\url{https://www.cosmos.esa.int/gaia}), processed by the {\it Gaia}
Data Processing and Analysis Consortium (DPAC,
\url{https://www.cosmos.esa.int/web/gaia/dpac/consortium}). Funding for the DPAC
has been provided by national institutions, in particular the institutions
participating in the {\it Gaia} Multilateral Agreement. This research has made use of the SIMBAD database,
operated at CDS, Strasbourg, France 
\end{acknowledgements}

\bibliographystyle{aa} 
\bibliography{biblio}

\appendix

\section{Details on light curve filtering steps}

\subsection{Bootstrap test for minimum light curve length}\label{appendix_bootstrap}

\begin{figure}[h!]
   \centering
      \includegraphics[width=1\linewidth]{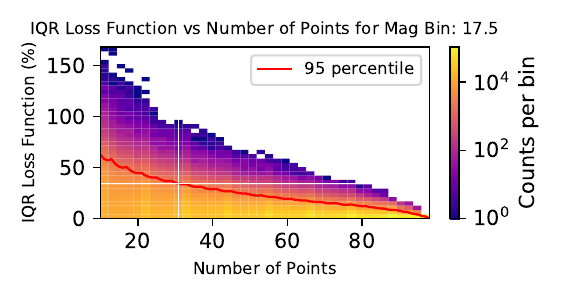}
      \caption{IQR variations as a function of the number of epochs for bootstrapped light curves in the magnitude bin 17.5 mag. The red line shows the 95$^{th}$ percentile of the distribution. The white cross indicates the threshold adopted for this magnitude bin.}
         \label{fig_iqr_loss_17.5}
   \end{figure}

We performed bootstrap tests to determine the minimum number of epochs required to ensure statistically robust variability indices. As a baseline, we selected light curves of GAPS sources that contained the highest number of epochs (typically \texttt{num\_selected\_g\_fov} between 70 and 100 observations). To account for variations of signal-to-noise as a function of observed magnitude, we selected 5 sources per bin with 0.5 mag width for $G$ between 6.5 and 20 mag. For each light curve selected, we computed the interquartile range (IQR) as a function of the number of epochs while randomly removing epochs until 10 epochs were left. To quantify the change introduced to the IQR by the removal of epochs, we computed a loss function as:

\begin{equation}
    L_{IQR}(i) = \frac{\left| IQR_{original} - IQR(i) \right|}{IQR_{original}} \times 100,
\end{equation}

\noindent where $IQR_{original}$ is the IQR of the original light curve and $IQR(i)$ the IQR at the $i^{th}$ step of epoch removal. The loss function represents a percentage difference between the original IQR and the IQR when points are removed. We repeated the process 10\,000 times per light curve to minimise biases introduced by specific epochs while randomly removing epochs. An example of the distribution of loss function values as a function of the number of epochs is shown in Fig.\ref{fig_iqr_loss_17.5} for the mag bin $G=17.5$m from where we inferred that a minimum of 31 epochs was required to keep the relative changes in the IQR under $1\sigma$ (34.1\%) for 95\% of the light curves in that magnitude bin (white cross).

\begin{figure}[h!bt]
   \centering
      \includegraphics[width=1\linewidth]{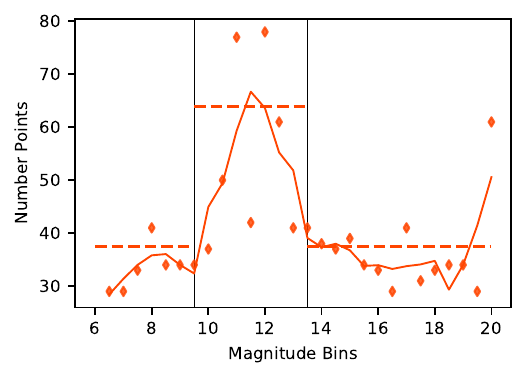}
      \caption{Minimum number of epochs needed to ensure the 95\% of sources have an IQR variation under 34.1\%, per mag bin. Diamonds show the raw values per mag bin. The solid line is a smoothed version. Vertical dotted lines show the 90 percentile of the points for a) mag bins between 9.5 and 13.5; b) other mag bins}
         \label{fig_threshold_40_percent}
   \end{figure}

Fig. \ref{fig_threshold_40_percent} shows the minimum number of epochs as a function of magnitude for all magnitude bins considered. There is a strong peak between 9.5 and 13.5 mags, meaning sources with this magnitude have an IQR changing more intensely, which is likely related to instrumental effects of Gaia around magnitude 12. To account for this feature, we applied a two-regime condition for the minimum number of observations and required at least 62 epochs for sources with magnitudes between 9.5 and 13.5 and 37 epochs for others.

\subsection{Spurious variability identification procedure}\label{app:spurioyuus_var}

As part of GDR3, \citet{Holl2023} provides a series of parameters calculated to facilitate the identification of spurious variability signals introduced by the complex scanning law of Gaia and varying object scanning angles. These parameters were retrieved for our sources from the \texttt{vari\_spurious\_signals} table through \cite{GaiaArchive}. In specific, we employed the pre-calculated Spearman correlation between photometric magnitudes and IPD GoF (Image Processing Data Goodness of Fit), \texttt{spearman\_corr\_ipd\_g\_fov}, and between magnitudes and their corrected flux excess factor \texttt{spearman\_corr\_exf\_g\_fov}. In both cases, a significant correlation suggests that scanning-law effects are likely present in the light curve.

\begin{figure}[htb]
   \centering
      \includegraphics[width=1\linewidth]{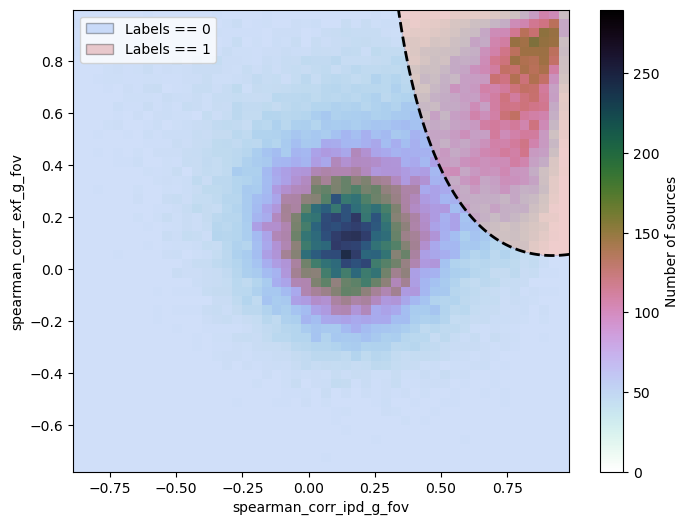}
      \caption{Distribution of Spearman correlations between the Gaia field-of-view photometric magnitudes and their corrected flux excess factor (y-axis) and their Image Processing Data Goodness of Fit (x-axis) for 79\,188 GDR3 YSOs. The distribution was split into separate Gaussian distributions, delimited by the blue and red shading.}
      \label{fig_corr_spurious_scan_angle}
   \end{figure}

Fig. \ref{fig_corr_spurious_scan_angle} shows the distribution of \texttt{spearman\_corr} \texttt{\_exf\_g\_fov} vs \texttt{spearman\_corr\_ipd\_g\_fov}, where the distribution is seen split into two bulks. We used a Gaussian Mixture model to isolate these two bulks, which allowed us to flag 22\,375 sources in the YSO sample as likely affected by spurious variability. 187 sources had no \texttt{spearman\_corr\_exf\_g\_fov} and were also discarded.

\subsection{Timescales in excess due to residual spurious variability}

\begin{figure}[htb]
   \centering
      \includegraphics[width=1\linewidth]{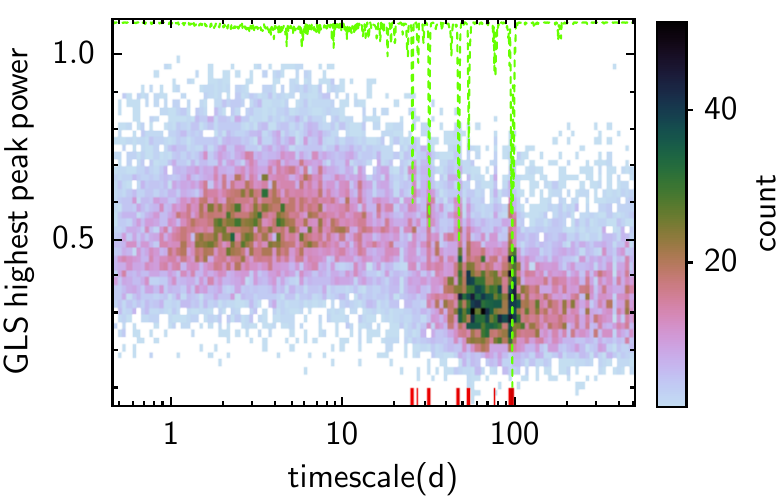}
      \caption{Density distribution of the power of the highest peak in the GLS periodograms as a function of their timescale for 34\,739 variable sources analysed in Sect.~\ref{gls_ts}, shown regardless of their FAL. The green dashed lines show the normalised frequency distribution of GLS timescales for sources affected by spurious variability introduced by the complex scanning law of Gaia and varying object scanning angles identified in Sect.~\ref{app:spurioyuus_var}. This distribution is renormalised here for visualisation purposes. The red rugged lines highlight timescales in excess in our sample due to residual spurious variability, and masked from our sample.}
      \label{spurious_residual}
   \end{figure}

In Fig.~\ref{spurious_residual}, we show a distribution of the power of the highest peak in the GLS periodogram as a function of the timescale for that peak (derived in Sect.~\ref{gls_ts}). An excess of sources detected at specific long timescales (e.g. $\sim$100, 50 and 25 d) is noticeable in this distribution. Fig.~\ref{spurious_residual} also shows a normalised frequency distribution of GLS timescales for sources affected by spurious variability identified in the previous section. Considering the timescales with at least 10\% frequency among sources with spurious variability points to a correspondence of these with the timescales in excess, suggesting a prevalence of sources with spurious variability in our sample. We have further inspected these sources in the context of \texttt{spearman\_corr\_ipd\_g\_fov} and \texttt{spearman\_corr\_exf\_g\_fov} parameters discussed in the previous section to address if a different strategy could reduce their incidence. For example, the requirement that rather than both, either of these parameters show stronger correlation or using a lower correlation threshold. However, a relevant population of residual spurious variability sources was found even among sources with correlation values close to zero. In particular, we note that the recurrence of residual spurious variability is minimal within the primary (highly) variable sample defined in Sect.~\ref{sec:selection_high_var}, but accentuated among sources in the secondary sample. Finally, we minimised this residual contamination by utilising the normalised frequency distribution of GLS timescales for sources affected by spurious variability identified in the previous section to mask the most frequent timescales among sources with spurious variability. 

\subsection{Procedure for outlier removal}\label{app:outlier}

Visual inspection of light curves exhibiting the highest peak-to-peak amplitude revealed that many sources had their variability and asymmetry analysis dominated by only one or two epochs beyond the $5\sigma$ level. In the context of the Gaia sparse light curves, we must address our capacity to judge whether these are spurious outliers or the results of the sparse sampling of a rapidly occurring variability event. For example, short-duration bursting features reported by \citet{Cody2017} typically span 10–20 days and typically show short duty cycles (time spent bursting). At certain latitudes, the Gaia scanning law can occasionally yield the occurrence of a dozen observed epochs over a couple of days, a phenomenon often referred to as cusps (see, for example, source $GDR3\,3336474807652948992$ in Fig.~\ref{fig:M_evolution}). However, if not coincident with cusps, the sparse sampling of Gaia typically includes only one epoch every 10-20 days, henceforth under-resolving and, likely, often missing short-duration bursts.

\begin{figure}[htb]
    \centering
    \includegraphics[width=0.99\linewidth]{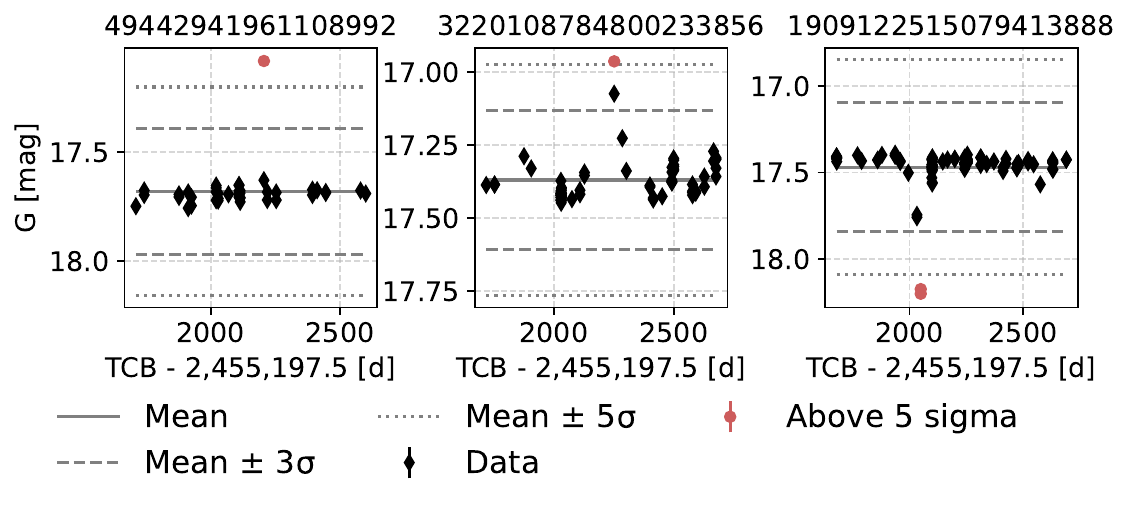}
    \caption{Examples of light curves with epochs beyond $5\sigma$ level. \emph{Left:} This epoch was not followed or preceded by a $3\sigma$ outlier, and it was thus identified as an outlier and removed from the light curve. \emph{Middle}: an example with a preceding epoch beyond the $3\sigma$ level considered as a reliable variability feature. \emph{Left}: Example with two successive epochs outside the $5\sigma$ level also considered as a reliable variability feature. }
    \label{fig_spurious_B}
\end{figure}

We identified 735 sources (2\% of the variable sample defined by Sect.~\ref{sec:spurious_outliers}) with at least one $5\sigma$ outlier and only 11 having two outliers at the same level. We further hypothesise that finding these outliers at a systematically bursting behaviour could support the idea that these are undersampled but reliable rapid burst variables. However, we find an overall homogeneous distribution of outlier behaviour with about $\sim 45\%$ flaring outliers versus $\sim55\%$ dimming outliers, thus denying our hypothesis. Including colour information could aid in further probing such events; nevertheless, colour variability was beyond the scope of the present study. We thus limit our approach to identifying and removing isolated outliers as follows. For each light curve with an epoch outside the $5\sigma$ level, we evaluated if this epoch was isolated by examining if the immediately close epochs were at least within the $3\sigma$ level while following the same bursting/dimming behaviour. If the epoch beyond the $5\sigma$ level failed this criterion, it was considered an outlier and removed from the light curve. Fig.~\ref{fig_spurious_B} shows examples where the first light curve shows an example with an outlier, and two others show light curves with epochs outside the $5\sigma$ level but identified as part of a reliable variability feature. 636 light curves had outliers removed with this criteria.

\section{Blind timescales}\label{app_SF_tau}

\begin{figure}
    \centering
    \includegraphics[width=0.98\linewidth]{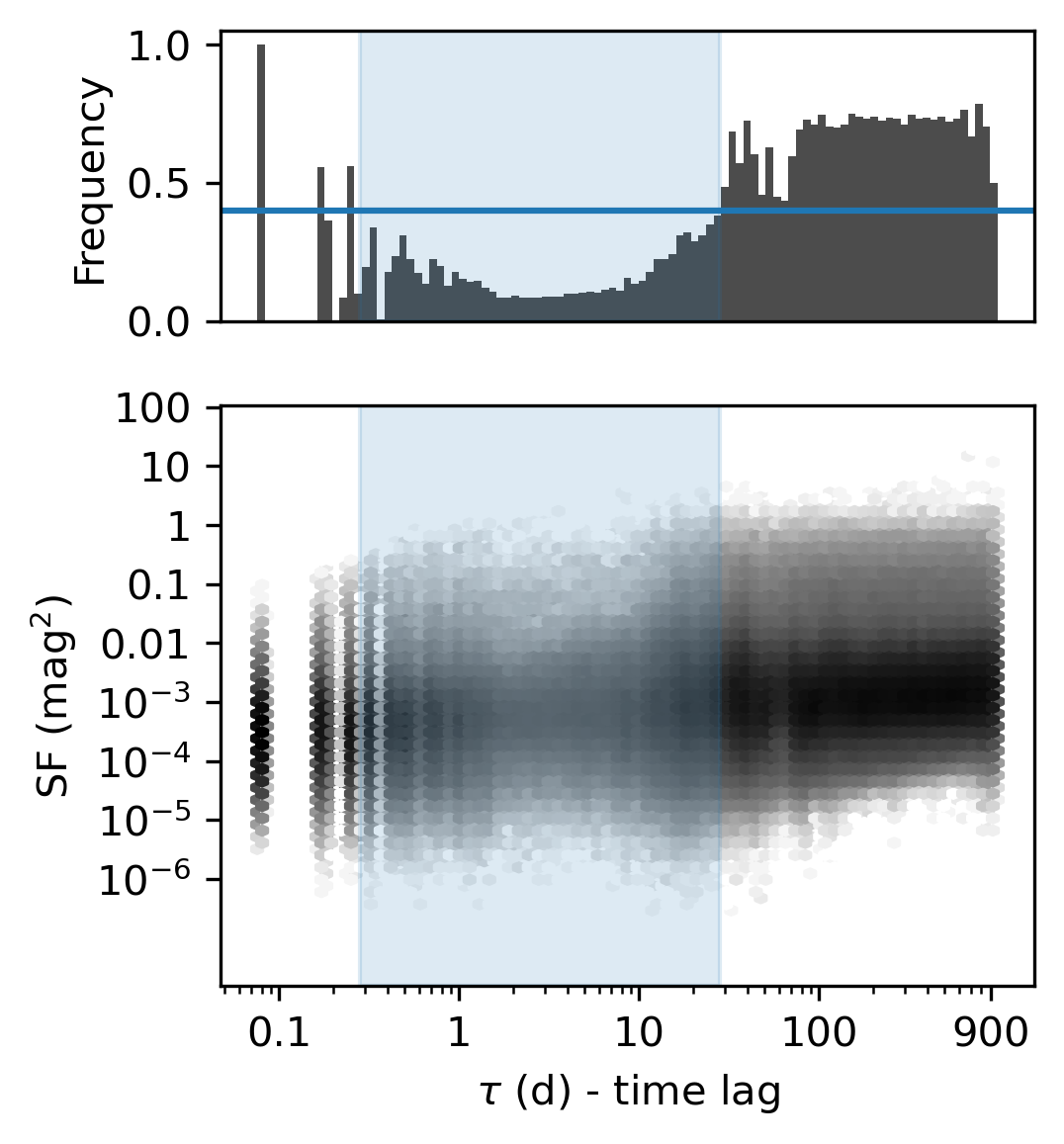}
      \caption{\emph{Bottom:} Structure function and time difference pairs considering all SF-points ($\tau$, SF) for all the 43\,840 sources with enough number of epochs (Sect.~\ref{sec:numb_epochs}). The blue region shows the depletion of pairs with $\tau$ between 0.3 and 28 days, introduced by limitation in the Gaia time-series sampling. \emph{Top:} Distribution of $\tau$ as a function of the frequency in which sources had a valid SF($\tau$). The blue horizontal line shows the 40\% limit used to infer the blind timescale range of Gaia.}
      \label{fig_sf_all}
   \end{figure}

In this Appendix, we investigate the range of timescales affected by the sparse time coverage of Gaia. To this end, we employed Eq.~\ref{sf_equation} to calculate structure functions for all 43\,840 sources with enough epochs in their light curves according to the procedure discussed in Sect.~\ref{sec:numb_epochs}. Fig.~\ref{fig_sf_all} shows the distribution of calculated $\tau$-SF($\tau$) pairs at a population level, where a lower density of points is clearly seen at $\tau$ values around $\sim1-10$ days. The frequency of $\tau$-SF($\tau$) pairs occurrence is also shown in the top panel of Fig.~\ref{fig_sf_all}.
We choose a conservative requirement of at least 40\% to define a blind-SF range of timescales in the range $\sim 0.3-28$ days. Nevertheless, we note that the issue of underpopulated SF is more critical for the timescale range of $\sim 0.5-10$ days where any given $\tau$-SF($\tau$) exists for less than 10\% of sources. GDR3 scanning law (Fig.~\ref{fig:sky_distributions}) implies that light curves will have varied sampling. Hence, the definition of this blind timescale range does not mean that timescales in this range cannot be measured with the SF method, but rather that SF will less frequently be well-sampled enough to detect such timescales. Based on the identification of this blind range,  Sect.~\ref{sec:numb_epochs} includes additional processing steps to identify if an aperiodic timescale in this range came from an adequately sampled SF. Finally, we also note that this sampling issue is only expected to affect the SF method with the underlying assumption of an aperiodic SF model. For GLS periodic timescales, different underlying assumptions of a sinusoidal periodic signal apply, and various previous studies based on GDR3 light curves have corroborated its efficiency in detecting timescales in this range \citep[e.g.][]{Distefano2022}.

\section{YSO properties}

\subsection{Ongoing accretion}\label{app:accretion}

We used the data described in Sect.~\ref{sec:YSOproperties} to classify sources as CTTS or WTTS based on their H$\alpha$ emission. For equivalent width measurements, we employed spectral type-dependent criteria by \citet{BarradoYNacascuesMartin2003} converted into Gaia $G_{BP}$-$G_{RP}$ colours using the \citet{PecautMamajek2013} empirical sequence. For full widths at 10\% of the peak flux measurements, we used the \citet{WhiteBasri2003} criterion to distinguish between CTTS and WTTS. For 64 sources, we used qualitative flags that report H$\alpha$ in emission \citep[from][]{Pettersson2014AandA...570A..30P}. When multiple H$\alpha$ measurements were available, we considered that one measurement indicating a CTTS classification was enough to grant the source a classification as such. Specifically for EW(H$\alpha$) measurements from ESP-ELS Gaia DR3 we have restricted our analysis to 7\,318 sources with the absolute value of reported EW(H$\alpha$) larger than at least 5 times the reported uncertainties. This decision was motivated by the limitations due to $G_{BP}$/$G_{RP}$ spectral low resolution \citep[$R\sim30-100$;][]{BPRPspectra2023AA...674A...2D} and its representation as Hermite function bases, where the identification of sources with large values of EW(H$\alpha$) are reliable, but where low accreting and non-accreting sources with EW(H$\alpha$) values closer to the stellar continuum may be confused within uncertainties. 

\subsection{Infrared indices}\label{app:alpha_ir}

To estimate $\alpha_\mathrm{IR}$ we followed a similar procedure as the one in the NEMESIS YSO catalogue \citep{roquette2024}: We used the Vizier Photometry viewer tool to retrieve for each source all previously published photometric data available at the Centre de données astronomiques de Strasbourg \citep[][]{Genova2000AAS..143....1G_CDS,Ochsenbein2000AAS..143...23O}, and { used all data available for} the wavelength range $2-24\,\mu$m to fit a linear model and measure $\alpha_\mathrm{IR}$, { with a requirement of at least two photometric data points 2$\mu m$ apart in this interval. Next, we adapted the standard classification scheme for YSOs based on their $\alpha_\mathrm{IR}$ \citep{Lada:87a, Andre1993, Greene1994} by grouping sources based on the dominant contributor to their SED: sources with significant envelope contribution have $\alpha_{IR} > -0.3$
(1\,378 sources), sources with a thick disc have $-0.3 < \alpha_{IR} \le -1.6$ (6\,427 sources), sources with thin discs have $-1.6 < \alpha_{IR} \le -2.5$ (8\,743 sources) and discless sources have $\alpha_{IR} < -2.5$, (11\,631 sources). }

\subsection{Spectral types}\label{app:SpT}

In GDR3, effective temperatures from the general stellar parameterizer from photometry\citep[GSP-Phot;][]{BPRPspectra2023AA...674A..27A} were available for 21\,112 of our sources. We combined these with the $T_\mathrm{eff}$-Spectral Type scale for PMS stars by \citet{PecautMamajek2013} to infer spectral types for our sample. For 7\,067 sources, effective temperatures were unavailable, and instead we used a $(G_{BP}-G_{RP})_0-$Spectral Type version of the \citet{PecautMamajek2013} scale. For this end, individual extinctions were inferred at the position and the GDR3 distance of sources using the Galactic 3D dust map by \citet{Edenhofer2024A&A...685A..82E} with the \texttt{dustmaps} Python package \citep{DustMaps2018JOSS....3..695M}.  \citet{Edenhofer2024A&A...685A..82E} provides a robust map for sources up to 1.25 kpc, with a secondary map based on less data available up to 2 kpc. For 84 sources located at larger distances, we adopted the 2 kpc extinction value as a bottom limit. Although informative, stellar parameters for young stars provided by GSP-Phot have a series of caveats \citep[see][for a in-depth discussion]{BPRPspectra2023AA...674A..27A}, including the use of time-averaged mean BP/RP spectra, which neglects YSO variability, and known degeneracy between effective temperature and extinction - expected to be especially relevant for sources in regions of enhanced line-of-sight extinction such as star-forming regions. Due to these known limitations, we did not attempt to derive sub-spectral types, and we stress that the reported spectral types are rough estimations aimed at a broad insight into the content of our YSO sample.

\section{Description of the data table}

\begin{table*}[h!]
    \caption{Overview of the table contents}\label{tab:description}
    \centering
    \renewcommand{\arraystretch}{1.35} 
    \begin{tabular}{lp{10.5cm}l} 
        \hline
        Name  & Description & Unit \\
        \hline
        source\_id & GDR3 source identifier & \\

        ra & Right ascension in ICRS at Gaia reference epoch & degree \\

        dec & Declination in ICRS at Gaia reference epoch & degree\\

        parallax & Absolute stellar parallax at Gaia reference epoch & mas \\

        phot\_g\_mean\_mag & G-band mean magnitude & mag \\

        num\_selected\_g\_fov  & Number of G FOV transits selected for variability analysis by Gaia & \\

        CTTS & If 0, H$\alpha$ observation exist but does not indicate ongoing accretion (WTTS). If > 0, at least one observation reports ongoing accretion (CTTS). If \texttt{null}, no available information. See App.~\ref{app:accretion}. & \\

        alpha\_ir  & $\alpha_\mathrm{IR}$ index. See App.~\ref{app:alpha_ir}. & \\
        
        { disc}\tablefootmark{a} & Infrared classification as in App.~\ref{app:alpha_ir}\\

        { nsedpoints} & number of points in the SED available for $\alpha_\mathrm{IR}$ derivation  & \\
        
        { min\_wl} & minimum wavelength available for  $\alpha_\mathrm{IR}$ derivation & $\mu m$\\

        { max\_wl} & maximum wavelength available for  $\alpha_\mathrm{IR}$ derivation & $\mu m$ \\
        
        timescale\_method  & GLS if the GLS periodogram was used to get the timescale (Sect.~\ref{gls_ts}). SF if the structure function was used (Sect.~\ref{sf_ts}). & \\

        timescale  & Dominant timescale, if not larger than half the light curve duration & days \\

        long\_timescale\_method & GLS if the GLS periodogram was used to get the timescale. SF if the structure function was used. Null if the dominant timescale is not long. & \\

        long\_timescale & Dominant timescale, if larger than half the light curve duration & days \\

        primary & True if the source is part of the primary sample. False if the source is part of the secondary sample. See Sect.~\ref{sec:selection_high_var}. & \\

        Q\_index & Q index value from Eq.~\ref{Q_eq} & \\

        M\_index & M index value from Eq.~\ref{M_eq} & \\

        amplitude & Peak-to-peak amplitude of the G-band light curve & mag \\

        morphological\_class & Morphological class inferred from Q-M indices. Can be P, EB, QPS, QPD, B, S, APD, L (Sect.~\ref{subsection_morpho_class}). & \\

        { SpT} & Spectral type estimation (App.\ref{app:SpT}) &  \\
        { SpT\_method}\tablefootmark{b} & Method for Spectral type estimation &  \\
        \hline
    \end{tabular}
    \tablefoot{ \tablefoottext{a}{\texttt{Discless}: sources without disc contribution detectable in their SED ($\alpha_{IR} \leq-2.5$); \texttt{Thin}: sources with detectable contribution of a thin disc ($-2.5\leq\alpha_{IR} \leq-1.6$); \texttt{Thick}: sources with detectable contribution of a thick disc ($-1.6\leq\alpha_{IR} \leq-0.3$) \texttt{Envelope}: sources with detectable contribution an envelope ($\alpha_{IR} \geq-0.3$)}; \tablefoottext{b}{\texttt{Teff->SpT}: Spectral Type estimated based on the comparison of the source GDR3 GSP-Phot temperature with an empirical sequence by Pecaut \& Mamajek (2013) ; \texttt{Bp-Rp->SpT} Spectral type estimated based comparison of its $(G_{BP}-G_{RP})_0$ with the empirical sequence. 
    }}
\end{table*}

Table~\ref{tab:description} provides an overview of the parameters used in this study, including source identifiers, astrometric and photometric properties, variability metrics, and classification results. 
The corresponding data table is available on Vizier.

\end{document}